\documentclass[journal,12pt,onecolumn]{IEEEtran}

\usepackage{setspace}
\ifCLASSOPTIONonecolumn \doublespace \fi

\usepackage{amsfonts}
\usepackage{amsmath}
\usepackage{graphicx}
\usepackage{cite}
\allowdisplaybreaks[1]

\graphicspath{{Figures/}} \DeclareGraphicsExtensions{.eps,.ps}

\title{Small Cell In-Band Wireless Backhaul in Massive MIMO Systems: A Cooperation of Next-Generation Techniques}

\author{Boyu~Li,~\IEEEmembership{Member,~IEEE,}~Dengkui~Zhu,~\IEEEmembership{Member,~IEEE,}~and~Ping~Liang
\thanks{
B. Li and D. Zhu are with RF DSP Inc., Irvine, CA 92606 USA (e-mail:
byli@rfdsp.com; dkzhu@rfdsp.com).}
\thanks{
P. Liang is with the University of California - Riverside, Riverside, CA
92521 USA and also with RF DSP Inc., Irvine, CA 92606 USA (e-mail:
liang@ee.ucr.edu).
}}

\begin{document}
\maketitle

\begin{abstract}
Massive Multiple-Input Multiple-Output (MIMO) systems, dense Small-Cells (SCs), and full duplex are three candidate techniques for next-generation communication systems. The cooperation of next-generation techniques could offer more benefits, e.g., SC in-band wireless backhaul in massive MIMO systems. In this paper, three strategies of SC in-band wireless backhaul in massive MIMO systems are introduced and compared, i.e., Complete Time-Division Duplex (CTDD), Zero-Division Duplex (ZDD), and ZDD with Interference Rejection (ZDD-IR). Simulation results demonstrate that SC in-band wireless backhaul has the potential to improve the throughput for massive MIMO systems. Specifically, among the three strategies, CTDD is the simplest one and could achieve decent throughput improvement. Depending on conditions, with the self-interference cancellation capability at SCs, ZDD could achieve better throughput than CTDD, even with residual self-interference. Moreover, ZDD-IR requires the additional interference rejection process at the BS compared to ZDD, but it could generally achieve better throughput than CTDD and ZDD. 
\end{abstract}


\section{Introduction} \label{sec:Introduction}

In the past few decades, significant development of information and communication technologies has been realized, tremendously improving our lives. Particularly, wireless communication systems have been playing a crucial role as the demand for wireless services has been constantly increasing. Nowadays, the latest standard for the Fourth Generation (4G) of mobile telecommunication, i.e., the Third Generation Partnership Project (3GPP) Long Term Evolution Advanced (LTE-A) standard \cite{3GPP_TS_36.201}, has been targeting downlink and uplink peak data rates of $1\mathrm{Gbps}$ and $500\mathrm{Mbps}$ respectively \cite{3GPP_TR_36.913}, which is already very challenging. Nonetheless, research interest has already been drawn on achieving substantially higher throughput than LTE-A, i.e., the Fifth Generation (5G) standard of mobile telecommunication. Obviously, the challenge is even bigger. Despite of the difficulties, some candidate solutions have been considered and under research. The three candidates that relate to this paper are briefly introduced below. 

The first approach is massive Multiple-Input Multiple-Output (MIMO) systems, which were firstly proposed in \cite{Marzetta_massive_MIMO_original}. A massive MIMO system considers scaling up conventional MIMO systems by possibly orders of magnitude, i.e., hundreds of antennas at a Base-Station (BS) simultaneously serve tens of User Equipments (UEs) in the same time-frequency resource. Such a system could provide tremendous advantages \cite{Marzetta_massive_MIMO_original, Rusek_massive_MIMO_overview, Hoydis_massive_MIMO, Larsson_massive_MIMO_overview}. With the capabilities of aggressive spatial multiplexing and great array gains, a massive MIMO system could achieve capacity increase and energy efficiency improvement simultaneously \cite{Marzetta_massive_MIMO_original, Rusek_massive_MIMO_overview, Larsson_massive_MIMO_overview, Hoydis_massive_MIMO}. In addition, it could be built with inexpensive and low-power components \cite{Larsson_massive_MIMO_overview}. Furthermore, it also has the potential to significantly reduce the latency on the air interface, simplify the multiple-access layer, as well as increase the robustness to both unintentional artificial interference and intended jamming \cite{Larsson_massive_MIMO_overview}. In general, massive MIMO systems are considered in Time-Division Duplex (TDD) mode, taking advantage of the channel reciprocity between the uplink and downlink \cite{Marzetta_massive_MIMO_original, Rusek_massive_MIMO_overview, Larsson_massive_MIMO_overview, Hoydis_massive_MIMO}, although such systems in Frequency-Division Duplex (FDD) mode have also been studied \cite{Adhikary_JSDM,Jiang_FDD_Massive_MIMO}.

The second one is based on high-density deployment of Small Cells (SCs). Although SCs are currently applied mainly for traffic offloading and indoor coverage, they have the potential to offer high capacity in a cost and energy efficient way, in both indoor and outdoor environments \cite{Hoydis_SCN, Andrews_femtocells}. In theory, the network capacity scales linearly with the cell density, so reducing the cell size could 
effectively improve network capacity. On the other hand, since a shorter distance results in less path loss, the total network transmitting power could be reduced, thus increasing the cost and energy efficiency.   

Another approach is the full-duplex technique based on self-interference cancellation. Instead of transmitting and receiving signals from separate times or frequencies as the currently employed half-duplex schemes, i.e., TDD and FDD, the full-duplex techniques have the ability to transmit and receive signals on the same frequency at the same time \cite{Radunovic_FD, Choi_FD, Knox_FD, Hua_FD_MIMO, Bharadia_FD}. As a result, improved capacity could be expected, as shown in \cite{Goyal_FD}. For the sake of distinction, the full-duplex scheme is called Zero-Division Duplex (ZDD) in this paper.   

Although the comparison between massive MIMO and dense SC networks has been carried out as in \cite{Liu_massive_MIMO_SC}, they are not necessarily competitors. In fact, they could be allies. For example, a network model was proposed in \cite{Hoydis_massive_MIMO_SC}, where each macro BS applies 
massive MIMO 
to support highly mobile UEs, while dense SCs are employed to support nominally mobile UEs. Although it is not the only way to incorporate massive MIMO and dense SCs, it does show that the cooperation of these two approaches could offer very high throughput. Moreover, it could be expected that higher throughput could be achieved if ZDD could be incorporated with them. 

Due to the potential advantages mentioned above, seeking applications with the cooperation of these three next-generation techniques is a very interesting issue. To that end, we consider a system where the massive MIMO technique is applied to offer in-band wireless backhaul for multiple SCs as a good application. The reasons are listed below.
\begin{enumerate}
\item Due to the fixed positions of SCs, the coherence time between each SC and a BS is relatively long. It is very desirable for massive MIMO techniques, especially when employing relatively complicated processes such as Zero-Forcing (ZF) \cite{Rusek_massive_MIMO_overview} and Interference Rejection (IR) \cite{Leost_IRC}. Using massive MIMO to provide wireless backhaul allows a high degree of spatial multiplexing, enabling the backhaul station to provide sufficient bandwidth to support multiple SCs using the same frequency resource, e.g., providing a $20\mathrm{MHz}$ backhaul band to many SCs using a single $20\mathrm{MHz}$ frequency band.

\begin{figure}[!t]
\centering \includegraphics[width = 0.5\linewidth]{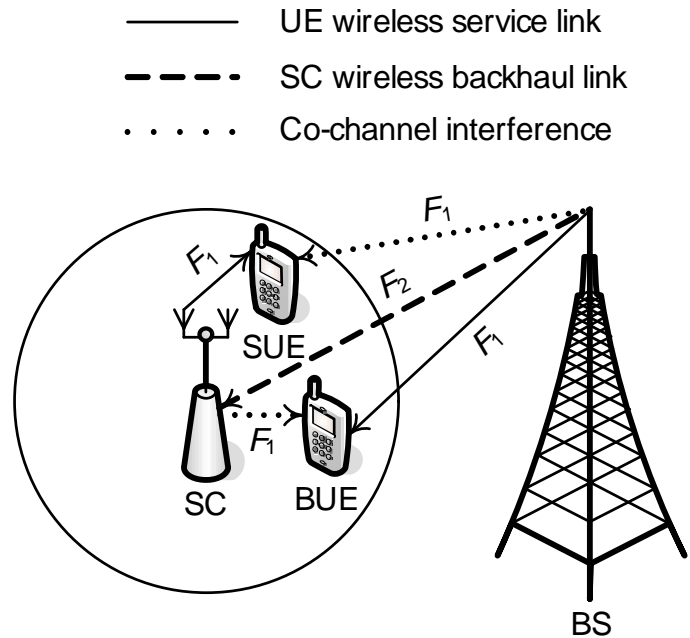}
\caption{A simple example of out-band wireless SC backhaul for the downlink.}
\label{fig:eg_outband}
\end{figure}
\begin{figure}[!t]
\centering \includegraphics[width = 0.5\linewidth]{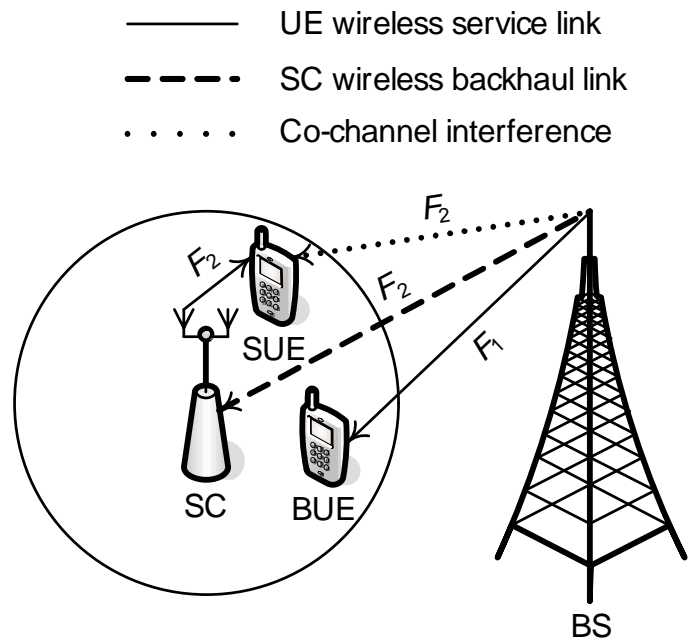}
\caption{A simple example of in-band wireless SC backhaul for the downlink.}
\label{fig:eg_inband}
\end{figure}

\item It is highly desirable to place SCs wherever they are needed, not only where wired backhaul is available. In this case, wireless backhaul \cite {SCF-049.04.02} is desired, especially for outdoor SCs where wired backhaul is hard to be offered. SC out-band wireless backhaul is operated in a separate frequency band, but the co-channel interference between a BS and each of its associated SCs still exists. For instance, Fig. \ref{fig:eg_outband} is a simple example of SC out-band wireless backhaul for the downlink consisting of a BS, a SC, and two UEs, where the frequency band $F_1$ is applied for UE service while the frequency band $F_2$ is used for SC wireless backhaul. In the figure, BUE and SUE denote the UEs associated with the BS and the SC respectively. The co-channel interference from the SC to the BUE as well as the co-channel interference from the BS to the SUE are shown in the figure. As a result, advanced schemes to deal with such co-channel interference in heterogeneous networks as enhanced Inter-Cell Interference Coordination (eICIC) \cite{Tran_LTE-A, R1-104968}, included in LTE-A, are needed. Note that eICIC requires additional exchanges of indicator-messages via the X2 interface or Operation and Management (O{\&}M), and the victim UEs are scheduled with reduced time or frequency resource or allocated with lower power to mitigate inter-cell interference \cite{Tran_LTE-A, R1-104968}. On the other hand, with the same two frequency bands, in-band SC wireless backhaul allows the associated SUEs to be operated in the same separate band as the SC in-band wireless backhaul, e.g., Fig. \ref{fig:eg_inband}. As a result, the co-channel interference from the SC to the BUE is completely removed. Hence, as long as the co-channel interference from the BS to the SUE can be properly addressed, which is investigated in this paper, the aforementioned advanced schemes such as eICIC are not required. Since the data transmission of SUEs uses the separate band, so it does not reduce the resources used for the data transmission of BUEs. Note that since a separate band is used for the data transmission of SUEs, it should be designed to be sufficient.

\item Although ZDD could be applied to MIMO systems, it is unlikely to be employed for massive MIMO due to the extremely high complexity, at least with current technology. Hence, a SC equipped with a few antennas is a good choice for the application of ZDD. With the capability of transmitting and receiving signals on the same frequency at the same time, ZDD could be applied for SCs employing in-band wireless backhaul.
\end{enumerate} 

Traditional millimeter wave wireless backhaul requires line-of-sight propagation \cite {SCF-049.04.02}. The goal of this paper is to investigate wireless backhaul that does not require line-of-sight. To the best of our knowledge, currently, no other literature considers using massive MIMO as in-band wireless backhaul for SCs. With in-band wireless backhaul, the BS-SC links and the SC-UE links are operated in the same band. It is not straightforward to deal with the interference introduced by the in-band wireless backhaul.  

In this paper, the potential throughput improvement introduced by SC in-band wireless backhaul in massive MIMO systems is discussed. Note that the paper focuses on analyzing the average achievable sum rate for backhaul transmission and data transmission with the allocated in-band bandwidth. At first, the basis of massive MIMO systems is briefly reviewed in Section \ref{sec:basis}. Then, three strategies are proposed in Section \ref{sec:strategies}. The first one is Complete TDD (CTDD), which further divides the communication between the BS and SCs and the data exchanges between SCs and their associated UEs in the time division. The second strategy is based on ZDD, which requires that SCs are capable to do self-interference cancellation thus could transmit and receive signals in the same time-frequency resource. The third one is called ZDD with IR (ZDD-IR). ZDD-IR is actually an enhanced version of ZDD, which has an additional requirement that the BS can apply the IR procedure. The average achievable sum rate values of downlink and uplink for the three strategies are derived. In Section \ref{sec:results}, simulation results of the three strategies are presented and compared to each other and basic massive MIMO systems, and the potential throughput gains of the three strategies are verified. Finally, conclusions are drawn in Section \ref{sec:conclusions}. Note that the major parameters of this paper are listed in Table \ref{table:parameters}.

\ifCLASSOPTIONonecolumn
\begin{table}[!t]
\renewcommand{\arraystretch}{1.3}
\caption{Major Parameters}
\centering
\begin{tabular}{|c|c|}
\hline
$B$ & System bandwidth \\
\hline
$\mathbf{H}_{\mathrm{b2s}},\mathbf{H}_{\mathrm{s2u}},\mathbf{H}_{\mathrm{b2u}}$ & Small-scale fading channels for the BS-SC, SC-UE, BS-UE links respectively \\
\hline
$\mathcal{H}_{\mathrm{b2s}},\mathcal{H}_{\mathrm{s2u}},\mathcal{H}_{\mathrm{b2u}}$ & Channels for the BS-SC, SC-UE, BS-UE links respectively \\
\hline
$K$ & Number of SCs or UEs. \\
\hline
$N_{\mathrm{bs}}$, $N_{\mathrm{sc}}$, and $N_{\mathrm{ue}}$ & Numbers of antennas for the BS, each SC, and each UE respectively \\
\hline
$N_{\mathrm{sct}}$ ($N_{\mathrm{scr}}$) & Number of transmitting (receiving) antennas of each ZDD-based SC \\
\hline
$P_{\mathrm{bs}}$, $P_{\mathrm{sc}}$, and $P_{\mathrm{ue}}$ & Maximal power values for the BS, each SC, and each UE respectively \\
\hline
$T$ & The time for a coherence time slot \\
\hline
$T_{\mathrm{dl}}$ ($T_{\mathrm{ul}}$) & Operation time in a coherence time slot for the downlink (uplink) \\
\hline
$\sigma^2_{\mathrm{bs}}$, $\sigma^2_{\mathrm{sc}}$, and $\sigma^2_{\mathrm{ue}}$ & Noise variances for the BS, each SC, and each UE respectively \\
\hline
\end{tabular}
\label{table:parameters}
\end{table}
\else
\vspace*{4pt}
\begin{table*}[!t]
\normalsize
\hrulefill
\renewcommand{\arraystretch}{1.3}
\caption{Major Parameters}
\centering
\begin{tabular}{|c|c|}
\hline
$B$ & System bandwidth \\
\hline
$\mathbf{H}_{\mathrm{b2s}},\mathbf{H}_{\mathrm{s2u}},\mathbf{H}_{\mathrm{b2u}}$ & Small-scale fading channels for the BS-SC, SC-UE, BS-UE links respectively \\
\hline
$\mathcal{H}_{\mathrm{b2s}},\mathcal{H}_{\mathrm{s2u}},\mathcal{H}_{\mathrm{b2u}}$ & Channels for the BS-SC, SC-UE, BS-UE links respectively \\
\hline
$K$ & Number of SCs or UEs. \\
\hline
$N_{\mathrm{bs}}$, $N_{\mathrm{sc}}$, and $N_{\mathrm{ue}}$ & Numbers of antennas for the BS, each SC, and each UE respectively \\
\hline
$N_{\mathrm{sct}}$ ($N_{\mathrm{scr}}$) & Number of transmitting (receiving) antennas of each ZDD-based SC \\
\hline
$P_{\mathrm{bs}}$, $P_{\mathrm{sc}}$, and $P_{\mathrm{ue}}$ & Maximal power values for the BS, each SC, and each UE respectively \\
\hline
$T$ & The time for a coherence time slot \\
\hline
$T_{\mathrm{dl}}$ ($T_{\mathrm{ul}}$) & Operation time in a coherence time slot for the downlink (uplink) \\
\hline
$\sigma^2_{\mathrm{bs}}$, $\sigma^2_{\mathrm{sc}}$, and $\sigma^2_{\mathrm{ue}}$ & Noise variances for the BS, each SC, and each UE respectively \\
\hline
\end{tabular}
\label{table:parameters}
\end{table*}
\fi 
\section{Massive MIMO Basis} \label{sec:basis}
In this section, the massive MIMO basis is briefly reviewed. Consider a TDD-based massive MIMO system where a BS is equipped with $N_{\mathrm{bs}}$ antennas and supports $K \ll N_{\mathrm{bs}}$ UEs in the same time-frequency resource. Assume that each UE has $N_{\mathrm{ue}}$ antennas and $N_{\mathrm{bs}} \gg KN_{\mathrm{ue}}$. Note that the assumption $N_{\mathrm{bs}} \gg KN_{\mathrm{ue}}$ is consistent with the general assumption of massive MIMO 
that the number of BS antennas is much higher than the spatial multiplexing streams \cite{Marzetta_massive_MIMO_original, Rusek_massive_MIMO_overview, Larsson_massive_MIMO_overview, Hoydis_massive_MIMO}. Then, the system channel is denoted by a $KN_{\mathrm{ue}} \times N_{\mathrm{bs}}$ matrix as
\begin{equation}
\mathcal{H}_{\mathrm{b2u}} = \left[\sqrt{a_{\mathrm{b2u},1}} \mathbf{H}^\mathrm{T}_{\mathrm{b2u},1} \, \cdots \, \sqrt{a_{\mathrm{b2u},K}} \mathbf{H}^\mathrm{T}_{\mathrm{b2u},K} \right]^{\mathrm{T}} 
\label{eq:channel_b2u}
\end{equation}
where $a_{\mathrm{b2u},k} < 1$ and the $N_{\mathrm{ue}} \times N_{\mathrm{bs}}$ matrix $\mathbf{H}_{\mathrm{b2u},k}$ denote the path loss and the small-scale fading channel coefficient matrix between the BS and the $k$th UE respectively with $k=1,\ldots,K$. The channel is assumed to be uncorrelated Rayleigh flat fading, i.e., the elements in $\mathbf{H}_{\mathrm{b2u},k}$ are independent and identically distributed (i.i.d.) zero-mean unit-variance complex Gaussian variables. Based on (\ref{eq:channel_b2u}), let
\begin{equation}
\mathbf{A}_{\mathrm{b2u}}=\mathrm{diag} \left[\sqrt{a_{\mathrm{b2u},1}}\mathbf{I}_{N_{\mathrm{ue}}} \, \cdots \, \sqrt{a_{\mathrm{b2u},K}}\mathbf{I}_{N_{\mathrm{ue}}} \right]
\label{eq:pathloss_b2u}
\end{equation} 
be a $KN_{\mathrm{ue}} \times KN_{\mathrm{ue}}$ diagonal matrix where $\mathbf{I}_{N}$ denotes the $N$-dimensional identity matrix, and let 
\begin{equation}
\mathbf{H}_{\mathrm{b2u}}=\left[\mathbf{H}^\mathrm{T}_{\mathrm{b2u},1} \, \cdots \, \mathbf{H}^\mathrm{T}_{\mathrm{b2u},K} \right]^{\mathrm{T}}.
\label{eq:channel_norm_b2u}
\end{equation}
Based on (\ref{eq:pathloss_b2u}) and (\ref{eq:channel_norm_b2u}), the relation (\ref{eq:channel_b2u}) can be rewritten as  
\begin{equation}
\mathcal{H}_{\mathrm{b2u}} = \mathbf{A}_{\mathrm{b2u}} \mathbf{H}_{\mathrm{b2u}}. 
\label{eq:channel_matrixform_b2u}
\end{equation}

For the downlink, ZF beamforming is employed, and the system input-output relation is
\begin{equation}
\mathbf{y}_{\mathrm{ue}} = \mathcal{H}_{\mathrm{b2u}} \mathcal{H}_{\mathrm{b2u}}^{\dagger} \mathbf{\Phi}^{\mathrm{dl}} \mathbf{x}_{\mathrm{bs}} + \mathbf{n}_{\mathrm{ue}} = \mathbf{\Phi}^{\mathrm{dl}} \mathbf{x}_{\mathrm{bs}} + \mathbf{n}_{\mathrm{ue}},
\label{eq:io_dl_b2u}
\end{equation}
where $(\cdot)^{\dagger}$ denotes the pseudo-inverse operation, the $(KN_{\mathrm{ue}})$-dimensional vectors $\mathbf{y}_{\mathrm{ue}}$, $\mathbf{x}_{\mathrm{bs}}$, and $\mathbf{n}_{\mathrm{ue}}$ denote the received signals at UEs, the transmitted signals at the BS with unit average power for each element, and the noise at UEs, respectively. The noise elements are assumed to be complex Additive White Gaussian Noise (AWGN) with variance $\sigma^2_{\mathrm{ue}}$. The $KN_{\mathrm{ue}} \times KN_{\mathrm{ue}}$ matrix $\mathbf{\Phi}^{\mathrm{dl}}$ is a diagonal matrix to satisfy the BS power constraints. Note that the short-term power constraint is assumed in this paper, i.e., the total power is allocated for only one channel realization. Let $P_{\mathrm{bs}}$ denote the maximum total BS power. In practice, each antenna has its own power amplifier, which means that each antenna has its own transmitting power constraint. Assume that each antenna has the same maximal power of $P_{\mathrm{bs}}/N_{\mathrm{bs}}$. For each antenna to satisfy its own power constraint, a linear scaling factor $\phi^{\mathrm{dl}}$ can be used so that at least one antenna works on full power and each UE has fair performance \cite{Shepard_argos}, where 
\begin{equation}
\phi^{\mathrm{dl}} = \sqrt{\frac{P_{\mathrm{bs}}}{N_{\mathrm{bs}}}} \left(\max \{ \|\mathcal{H}^{\dagger}_{\mathrm{b2u},i} \| \}_{i=1}^{N_{\mathrm{bs}}} \right)^{-1},
\label{eq:constraint_scalar_b2u}
\end{equation}
where $\mathcal{H}^{\dagger}_{\mathrm{b2u},i}$ is the $i$th row of $\mathcal{H}^{\dagger}_{\mathrm{b2u}}$ and $\|\cdot\|$ denotes the Euclidean norm. Then, in this case, $\mathbf{\Phi}^{\mathrm{dl}}$ reduces to the global scalar $\phi^{\mathrm{dl}}$. Hence, with the condition (\ref{eq:constraint_scalar_b2u}), the post-processing Signal-to-Interference-plus-Noise-Ratio (SINR) for each UE is
\begin{equation}
\gamma^{\mathrm{dl}} = \frac{\left(\phi^{\mathrm{dl}}\right)^2}{B\sigma^2_{\mathrm{ue}}}, 
\label{eq:sinr_dl_b2u}
\end{equation}
where $B$ is the system bandwidth. Then, for the $k$th UE, the average achievable throughput in $\mathrm{bit/s}$ is 
\begin{equation}
c^{\mathrm{dl},k} = \mathsf{E} \left\lbrace B N_{\mathrm{ue}} \log_2\left( \gamma^{\mathrm{dl}} +1 \right) \right\rbrace,
\label{eq:capacity_dl_b2u_wt}
\end{equation}
with $k=1,\ldots, K$. Let $T$ be the coherence time, and define $T_{\mathrm{dl}} \in [0, T]$ as the downlink operation time in a coherence time slot. Then, the average achievable throughput removing resources not used for transmission for the $k$th UE is 
\begin{equation}
C^{\mathrm{dl},k} = \left(\frac{T_{\mathrm{dl}}}{T}\right) c^{\mathrm{dl},k}.
\label{eq:capacity_dl_b2u}
\end{equation}
Based on (\ref{eq:constraint_scalar_b2u})-(\ref{eq:capacity_dl_b2u}), the term $(\max \{ \|\mathcal{H}^{\dagger}_{\mathrm{b2u},i} \| \}_{i=1}^{N_{\mathrm{bs}}})^{-1}$ in (\ref{eq:constraint_scalar_b2u}) is a key factor for the downlink average achievable throughput. Note that $\mathcal{H}_{\mathrm{b2u}}^{\dagger}= \mathbf{H}^{\dagger}_{\mathrm{b2u}} \mathbf{A}^{\dagger}_{\mathrm{b2u}}$ and $\mathbf{A}^{\dagger}_{\mathrm{b2u}}$ is a diagonal matrix. Since a worse path-loss matrix $\mathbf{A}_{\mathrm{b2u}}$ results in a better $\mathbf{A}^{\dagger}_{\mathrm{b2u}}$, then the term $(\max \{ \|\mathcal{H}^{\dagger}_{\mathrm{b2u},i} \| \}_{i=1}^{N_{\mathrm{bs}}})^{-1}$ decreases. As a result, $C^{\mathrm{dl},k}$ reduces. According to (\ref{eq:constraint_scalar_b2u})-(\ref{eq:capacity_dl_b2u}), the downlink average achievable sum rate is  
\begin{equation}
C^{\mathrm{dl}} = \sum_{k=1}^{K} C^{\mathrm{dl},k} = K C^{\mathrm{dl},k}.
\label{eq:sum_capacity_dl_b2u}
\end{equation}

As for the uplink, ZF decoding is applied, thus the system input-output relation is
\ifCLASSOPTIONonecolumn
\begin{equation}
\mathbf{y}_{\mathrm{bs}} = (\mathcal{H}_{\mathrm{b2u}}^{\dagger})^\mathrm{T} \mathcal{H}^\mathrm{T}_{\mathrm{b2u}} \mathbf{\Phi}^{\mathrm{ul}} \mathbf{x}_{\mathrm{ue}} + (\mathcal{H}_{\mathrm{b2u}}^{\dagger})^\mathrm{T} \mathbf{n}_{\mathrm{bs}} = \mathbf{\Phi}^{\mathrm{ul}} \mathbf{x}_{\mathrm{ue}} + (\mathcal{H}_{\mathrm{b2u}}^{\dagger})^\mathrm{T} \mathbf{n}_{\mathrm{bs}},
\label{eq:io_ul_u2b}
\end{equation}
\else
\begin{align}
\mathbf{y}_{\mathrm{bs}} & = (\mathcal{H}_{\mathrm{b2u}}^{\dagger})^\mathrm{T} \mathcal{H}^\mathrm{T}_{\mathrm{b2u}} \mathbf{\Phi}^{\mathrm{ul}} \mathbf{x}_{\mathrm{ue}} + (\mathcal{H}_{\mathrm{b2u}}^{\dagger})^\mathrm{T} \mathbf{n}_{\mathrm{bs}} \nonumber \\ & = \mathbf{\Phi}^{\mathrm{ul}} \mathbf{x}_{\mathrm{ue}} + (\mathcal{H}_{\mathrm{b2u}}^{\dagger})^\mathrm{T} \mathbf{n}_{\mathrm{bs}},
\label{eq:io_ul_u2b}
\end{align}
\fi
where the $(KN_{\mathrm{ue}})$-dimensional vectors $\mathbf{y}_{\mathrm{bs}}$ and $\mathbf{x}_{\mathrm{ue}}$ are the received signals at the BS and the transmitted signals at UEs with unit average power for each element, respectively, and the $N_{\mathrm{bs}}$-dimensional vector $\mathbf{n}_{\mathrm{bs}}$ denotes the noise at the BS whose elements are assumed to be complex AWGN with variance $\sigma^2_{\mathrm{bs}}$. The $KN_{\mathrm{ue}} \times KN_{\mathrm{ue}}$ matrix $\mathbf{\Phi}^{\mathrm{ul}}$ is a diagonal matrix to satisfy the power constraints of UEs. In this paper, equal power allocation is applied to achieve fairness for the data streams of each UE. Note that since each UE antenna is used to transmit an independent modulated data stream, applying equal power allocation automatically satisfies per-antenna power constraints, which is the same as the downlink. As a result, $\mathbf{\Phi}^{\mathrm{ul}}$ reduces to a global scalar $\phi^{\mathrm{ul}}$ as
\begin{equation}
\phi^{\mathrm{ul}} = \sqrt{\frac{P_{\mathrm{ue}}}{N_{\mathrm{\mathrm{ue}}}}},
\label{eq:constraint_scalar_u2b}
\end{equation}
where $P_{\mathrm{ue}}$ denotes the maximal power of each UE. Therefore, the post-processing SINR for the $i$th stream of the $k$th UE is  
\begin{equation}
\gamma^{\mathrm{ul}}\left(i,k\right) = \frac{\left(\phi^{\mathrm{ul}}\right)^2}{B\sigma^2_{\mathrm{bs}}} \|(\mathcal{H}^{\dagger}_{\mathrm{b2u}})^\mathrm{T}_j \|^{-2}, 
\label{eq:sinr_ul_u2b}
\end{equation}
where $(\mathcal{H}^{\dagger}_{\mathrm{b2u}})^\mathrm{T}_j$ is the $j$th column of $\mathcal{H}^{\dagger}_{\mathrm{b2u}}$, and $j=i+(k-1)N_{\mathrm{ue}}$ with $i=1,\ldots, N_{\mathrm{ue}}$, and $k=1,\ldots, K$. As a result, for the $k$th UE, the average achievable throughput in $\mathrm{bit/s}$ is  
\begin{equation}
c^{\mathrm{ul},k} = \mathsf{E} \left\lbrace B \sum_{i=1}^{N_{\mathrm{ue}}}\log_2\left[ \gamma^{\mathrm{ul}}\left(i,k\right)  +1 \right] \right\rbrace.
\label{eq:capacity_ul_u2b_wt}
\end{equation}
Let $T_{\mathrm{ul}} = T - T_{\mathrm{dl}}$ be the uplink operation time in a coherence time slot, then the average achievable throughput removing resources not used for transmission for the $k$th UE is
\begin{equation}
C^{\mathrm{ul},k} = \left(\frac{T_{\mathrm{ul}}}{T}\right) c^{\mathrm{ul},k}.
\label{eq:capacity_ul_u2b}
\end{equation}
According to (\ref{eq:constraint_scalar_u2b})-(\ref{eq:capacity_ul_u2b}), the term $\|(\mathcal{H}^{\dagger}_{\mathrm{b2u}})^\mathrm{T}_j \|^{-2}$ in (\ref{eq:constraint_scalar_u2b}) is a key factor for the uplink average achievable throughput. Note that $\mathcal{H}_{\mathrm{b2u}}^{\dagger}= \mathbf{H}^{\dagger}_{\mathrm{b2u}} \mathbf{A}^{\dagger}_{\mathrm{b2u}}$. Because a worse path-loss matrix $\mathbf{A}_{\mathrm{b2u}}$ causes a better $\mathbf{A}^{\dagger}_{\mathrm{b2u}}$, hence the term $\|(\mathcal{H}^{\dagger}_{\mathrm{b2u}})^\mathrm{T}_j \|^{-2}$ reduces, which results in a lower $C^{\mathrm{ul},k}$. Based on (\ref{eq:constraint_scalar_u2b})-(\ref{eq:capacity_ul_u2b}), the uplink average achievable sum rate is  
\begin{equation}
C^{\mathrm{ul}} = \sum_{k=1}^{K} C^{\mathrm{ul},k}.
\label{eq:sum_capacity_ul_u2b}
\end{equation}
\section{Strategies of SC In-Band Wireless Backhaul in Massive MIMO systems}\label{sec:strategies} 

As shown in Section \ref{sec:basis}, when the path losses of UEs are severe, massive MIMO systems could not 
provide sufficient throughput to each UE. When SCs with much smaller coverage are introduced into the system, the throughput could be significantly improved because it suffers much less path losses. In this paper, it is assumed that each SC is associated to only one UE in the same time-frequency resource, e.g., a Resource Block (RB) in LTE/LTE-A systems. For the sake of simplicity, each UE is assumed to be allocated all time-frequency resources of its associated SC, which can be easily generalized to the case of multiple UEs as discussed in Section \ref{subsec:discussions}. In addition, SCs are assumed to be carefully located so that inter-SC interference can be neglected. For wired backhaul and out-band wireless backhaul, the communication between the BS and SCs is separated from the data exchange between SCs and their related UEs. Then, as long as the throughput between the BS and each SC is good enough, the whole system would work. On the other hand, in the case of in-band wireless backhaul, the challenge is that the data exchange between the BS and SCs is now not independent of the communication between SCs and their associated UEs. Three strategies are discussed in this section below.

\subsection{CTDD} \label{subsec:strategies_ctdd}
Since the considered massive MIMO system is based on TDD, a simple way is to further separate the communication between the BS and SCs, and the data exchange between SCs and their related UEs, in the time division. 

For the data exchange between the BS and SCs, as each SC can be considered as a UE, the results (\ref{eq:constraint_scalar_b2u})-(\ref{eq:capacity_dl_b2u_wt}) can be applied to the downlink from the BS to each SC, and the results (\ref{eq:constraint_scalar_u2b})-(\ref{eq:capacity_ul_u2b_wt}) can be used for the uplink from each SC to the BS. Assume that each SC is equipped with $N_{\mathrm{sc}}$ antennas and $KN_{\mathrm{sc}} \ll N_{\mathrm{bs}}$. Note that the assumption $N_{\mathrm{bs}} \gg KN_{\mathrm{sc}}$ is consistent with the general assumption of massive MIMO systems that the number of BS antennas is much higher than the spatial multiplexing streams \cite{Marzetta_massive_MIMO_original, Rusek_massive_MIMO_overview, Larsson_massive_MIMO_overview, Hoydis_massive_MIMO}. Similarly to Section \ref{sec:basis}, define the $KN_{\mathrm{sc}} \times N_{\mathrm{bs}}$ channel matrix as 
\begin{equation}
\mathcal{H}_{\mathrm{b2s}} = \mathbf{A}_{\mathrm{b2s}} \mathbf{H}_{\mathrm{b2s}},
\label{eq:channel_matrixform_b2s}
\end{equation}
where 
\begin{equation}
\mathbf{A}_{\mathrm{b2s}}=\mathrm{diag} \left[\sqrt{a_{\mathrm{b2s},1}}\mathbf{I}_{N_{\mathrm{sc}}} \, \cdots \, \sqrt{a_{\mathrm{b2s},K}}\mathbf{I}_{N_{\mathrm{sc}}} \right]
\label{eq:pathloss_b2s}
\end{equation} 
is a $KN_{\mathrm{sc}} \times KN_{\mathrm{sc}}$ diagonal matrix whose element $a_{\mathrm{b2s},k} < 1$ denotes the path loss between the BS and the $k$th SC with $k=1,\ldots,K$, and  
\begin{equation}
\mathbf{H}_{\mathrm{b2s}}=\left[\mathbf{H}^\mathrm{T}_{\mathrm{b2s},1} \, \cdots \, \mathbf{H}^\mathrm{T}_{\mathrm{b2s},K} \right]^{\mathrm{T}}
\label{eq:channel_norm_b2s}
\end{equation}
with the $N_{\mathrm{sc}} \times N_{\mathrm{bs}}$ matrix $\mathbf{H}_{\mathrm{b2s},k}$ being the small-scale fading channel matrix between the BS and the $k$th SC in uncorrelated Rayleigh flat fading. Then, from the BS to the $k$th SC, the downlink average achievable throughput in $\mathrm{bit/s}$ is 
\ifCLASSOPTIONonecolumn
\begin{equation}
c^k_{\mathrm{b2s},1}\left( N_{\mathrm{sc}} \right) = \mathsf{E} \left\lbrace B N_{\mathrm{sc}} \log_2\left[ \frac{P_{\mathrm{bs}} }{ BN_{\mathrm{bs}}\sigma^2_{\mathrm{sc}} } \left(\max \{ \|\mathcal{H}^{\dagger}_{\mathrm{b2s},i} \| \}_{i=1}^{N_{\mathrm{bs}}} \right)^{-2} +1 \right] \right\rbrace,
\label{eq:capacity_dl_b2s_wt}
\end{equation}
\else
\begin{align}
c^k_{\mathrm{b2s},1}\left( N_{\mathrm{sc}} \right) & = \mathsf{E} \left\lbrace B N_{\mathrm{sc}} \log_2\left[ \frac{P_{\mathrm{bs}}}{ BN_{\mathrm{bs}}\sigma^2_{\mathrm{sc}} } \right. \right. \nonumber \\ & \left. \left. \,\,\, \times \left(\max \{ \|\mathcal{H}^{\dagger}_{\mathrm{b2s},i} \| \}_{i=1}^{N_{\mathrm{bs}}} \right)^{-2} +1 \right] \right\rbrace,
\label{eq:capacity_dl_b2s_wt}
\end{align}
\fi
where $\sigma^2_{\mathrm{sc}}$ denotes the complex AWGN noise variance for each SC. Let $T_{\mathrm{b2s}} \in [0, T_{\mathrm{dl}}]$ denote the transmission time from the BS to SCs, then the downlink average achievable throughput from the BS to the $k$th SC removing resources not used for transmission can be written as
\begin{equation}
C^k_{\mathrm{b2s},1}\left( N_{\mathrm{sc}} \right) = \left(\frac{T_{\mathrm{b2s}}}{T}\right) c^k_{\mathrm{b2s},1}\left( N_{\mathrm{sc}} \right).
\label{eq:capacity_dl_b2s}
\end{equation}
On the other hand, from the $k$th SC to the BS, the uplink average achievable throughput in $\mathrm{bit/s}$ is  
\ifCLASSOPTIONonecolumn
\begin{equation}
c^k_{\mathrm{s2b},1}\left( N_{\mathrm{sc}} \right) = \mathsf{E} \left\lbrace B \sum_{i=1}^{N_{\mathrm{sc}}}\log_2\left[ \frac{P_{\mathrm{sc}} }{ BN_{\mathrm{sc}}\sigma^2_{\mathrm{bs}}} \|(\mathcal{H}^{\dagger}_{b2s})^\mathrm{T}_j \|^{-2} +1 \right] \right\rbrace,
\label{eq:capacity_ul_s2b_wt}
\end{equation}
\else
\begin{align}
c^k_{\mathrm{s2b},1}\left( N_{\mathrm{sc}} \right) & = \mathsf{E} \left\lbrace B \sum_{i=1}^{N_{\mathrm{sc}}}\log_2\left[ \frac{P_{\mathrm{sc}} }{ BN_{\mathrm{sc}}\sigma^2_{\mathrm{bs}}} \right. \right. \nonumber \\ & \left. \left. \,\,\, \times \|(\mathcal{H}^{\dagger}_{\mathrm{b2s}})^\mathrm{T}_j \|^{-2} +1 \right] \right\rbrace,
\label{eq:capacity_ul_s2b_wt}
\end{align}
\fi
where $j=i+(k-1)N_{\mathrm{sc}}$ with $i=1,\ldots, N_{\mathrm{sc}}$, and $P_{\mathrm{sc}}$ is the maximal power of each SC.
Define $T_{\mathrm{s2b}} \in [0, T_{\mathrm{ul}}]$ as the transmission time from SCs to the BS, then the uplink average achievable throughput from the $k$th SC to the BS removing resources not used for transmission is
\begin{equation}
C^k_{\mathrm{s2b},1}\left( N_{\mathrm{sc}} \right) = \left(\frac{T_{\mathrm{s2b}}}{T}\right) c^k_{\mathrm{s2b},1}\left( N_{\mathrm{sc}} \right).
\label{eq:capacity_ul_s2b}
\end{equation}

As for the communication between each SC and its associated UE, it is in fact a general point-to-point MIMO system. Let the $N_{\mathrm{ue}} \times N_{\mathrm{sc}}$ matrix
\begin{equation}
\mathcal{H}_{\mathrm{s2u},k} = \sqrt{a_{\mathrm{s2u},k}} \mathbf{H}_{\mathrm{s2u},k}
\label{eq:channel_norm_s2u}
\end{equation}
denote the channel between the $k$th SC and its related UE, where $a_{\mathrm{s2u},k}<1$ is the path loss, and the $N_{\mathrm{ue}} \times N_{\mathrm{sc}}$ matrix $\mathbf{H}_{\mathrm{s2u},k}$ is the small-scale fading channel coefficient matrix in uncorrelated Rayleigh flat fading. Assume that $\mathcal{H}_{\mathrm{s2u},k}$ is known by both the $k$th SC and UE, then the optimal average achievable throughout is given in \cite{Goldsmith_WC_Ch10}. Specifically, the Singular Value Decomposition (SVD) of $\mathbf{H}_{\mathrm{s2u},k}$ is 
\begin{equation}
\mathbf{H}_{\mathrm{s2u},k} = \mathbf{U} \mathbf{\Lambda} \mathbf{V}^{H},
\label{eq:svd_s2u}
\end{equation}
where the $N_{\mathrm{ue}} \times N_{\mathrm{ue}}$ matrix $\mathbf{U}$ and the $N_{\mathrm{sc}} \times N_{\mathrm{sc}}$ matrix $\mathbf{V}$ are unitary. The $N_{\mathrm{ue}} \times N_{\mathrm{sc}}$ matrix $\mathbf{\Lambda}$ is rectangular diagonal whose $s$th diagonal element $\lambda_s \in \mathbb{R}^+$ is a singular value of $\mathbf{H}_{\mathrm{s2u},k}$ in decreasing order with $s=1,\ldots,\min\{N_{\mathrm{ue}}, N_{\mathrm{sc}}\}$, where $\mathbb{R}^+$ denotes the set of positive real numbers. When $S \in \{1,\ldots,\min\{N_{\mathrm{ue}}, N_{\mathrm{sc}}\}\}$ parallel streams are transmitted simultaneously, the first $S$ columns of $\mathbf{U}$ and $\mathbf{V}$, i.e., $\mathbf{U}_S$ and $\mathbf{V}_S$ are chosen as beamforming matrices at the UE and SC, respectively. Then, the optimal power allocation method is the water-filling power allocation as
\begin{equation}
\frac{P_s}{P} = \left\lbrace 
\begin{array}{cc}
\frac{1}{\gamma_0} - \frac{1}{\gamma_s}, & \gamma_s \geq \gamma_0, \\
0, & \gamma_s < \gamma_0,
\end{array}
\right.
\label{eq:wf_s2u}
\end{equation}
where $P \in \{P_{\mathrm{sc}},P_{\mathrm{ue}}\}$ and $P_s \in \{P_{\mathrm{sc},s}( k ),P_{\mathrm{ue},s}( k )\}$ are the total transmitting power and the transmitting power of the $s$th stream for the $k$th SC-UE pair respectively, the term $\gamma_s$ is defined based on \cite{Goldsmith_WC_Ch10} as 
\begin{equation}
\gamma_s = \frac{a_{\mathrm{s2u},k} \lambda^2_s P_s}{B\sigma^2} \in \left\lbrace\gamma_{\mathrm{s2u},s}\left( k \right),\gamma_{\mathrm{u2s},s}\left( k \right) \right\rbrace
\label{eq:wf_gamma}
\end{equation}
with $\sigma^2 \in \{\sigma^2_{\mathrm{ue}},\sigma^2_{\mathrm{sc}}\}$ being the complex AWGN noise variance, and $\gamma_0 \in \{\gamma_{\mathrm{s2u},0}(k),\gamma_{\mathrm{u2s},0}(k) \}$ is the cutoff value. Assume that $S \in \{S_{\mathrm{s2u}}(k), S_{\mathrm{u2s}}(k) \}$ is the number of parallel streams supported by the water-filling power allocation (\ref{eq:wf_s2u}) \cite{Goldsmith_WC_Ch10}. Then, from the $k$th SC to its associated UE, the average achievable throughput in $\mathrm{bit/s}$ is    
\begin{equation}
c^k_{\mathrm{s2u},1}\left( N_{\mathrm{sc}} \right) = \mathsf{E} \left\lbrace B \sum^{S_{\mathrm{s2u}}\left( k \right)}_{s=1} \log_2\left[ \frac{\gamma_{\mathrm{s2u},s}\left( k \right)}{\gamma_{\mathrm{s2u},0}\left( k \right)} \right] \right\rbrace.
\label{eq:capacity_dl_s2u_wt}
\end{equation}
Define $T_{\mathrm{s2u}} = T_{\mathrm{dl}} - T_{\mathrm{b2s}}$ as the transmission time from SCs to their related UEs, then the average achievable throughput from the $k$th SC to its associated UE removing resources not used for transmission is
\begin{equation}
C^k_{\mathrm{s2u},1}\left( N_{\mathrm{sc}} \right) = \left(\frac{T_{\mathrm{s2u}}}{T}\right) c^k_{\mathrm{s2u},1}\left( N_{\mathrm{sc}} \right).
\label{eq:capacity_dl_s2u}
\end{equation}
On the other hand, from the $k$th UE to its related SC, the average achievable throughput in $\mathrm{bit/s}$ is 
\begin{equation}
c^k_{\mathrm{u2s},1} \left( N_{\mathrm{sc}} \right) = \mathsf{E} \left\lbrace B \sum^{S_{\mathrm{u2s}}\left( k \right)}_{s=1} \log_2\left[ \frac{\gamma_{\mathrm{u2s},s}\left( k \right)}{\gamma_{\mathrm{u2s},0}\left( k \right)} \right] \right\rbrace.
\label{eq:capacity_ul_u2s_wt}
\end{equation}
Similarly, let $T_{\mathrm{u2s}} = T_{\mathrm{ul}} - T_{\mathrm{s2b}}$ be the transmission time from UEs to their associated SCs, then the average achievable throughput from the $k$th UE to its related SC removing resources not used for transmission is 
\begin{equation}
C^k_{\mathrm{u2s},1} \left( N_{\mathrm{sc}} \right) = \left(\frac{T_{\mathrm{u2s}}}{T}\right) c^k_{\mathrm{u2s},1} \left( N_{\mathrm{sc}} \right).
\label{eq:capacity_ul_u2s}
\end{equation}

According to (\ref{eq:capacity_dl_b2s_wt}), (\ref{eq:capacity_dl_b2s}), (\ref{eq:capacity_dl_s2u_wt}), and (\ref{eq:capacity_dl_s2u}), the downlink average achievable sum rate is 
\begin{equation}
C^{\mathrm{dl}}_{1}\left( N_{\mathrm{sc}} \right) = \sum_{k=1}^K \min \left\lbrace C^k_{\mathrm{b2s},1}\left( N_{\mathrm{sc}} \right), C^k_{\mathrm{s2u},1}\left( N_{\mathrm{sc}} \right) \right\rbrace.
\label{eq:capacity_dl_b2u_ctdd}
\end{equation}
Similarly, based on (\ref{eq:capacity_ul_s2b_wt}), (\ref{eq:capacity_ul_s2b}), (\ref{eq:capacity_ul_u2s_wt}), and (\ref{eq:capacity_ul_u2s}), the uplink average achievable sum rate is  
\begin{equation}
C^{\mathrm{ul}}_{1}\left( N_{\mathrm{sc}} \right) = \sum_{k=1}^K \min  \left\lbrace C^k_{\mathrm{u2s},1}\left( N_{\mathrm{sc}} \right), C^k_{\mathrm{s2b},1}\left( N_{\mathrm{sc}} \right) \right\rbrace.
\label{eq:capacity_ul_u2b_ctdd}
\end{equation}

For CTDD, the time allocation for each of the four aforementioned communication parts can be easily controlled. For a certain time duration, the best allocation strategy for a BS-UE link satisfies that the throughput between the BS and the $k$th SC equals to the throughput between the $k$th SC and its associated UE, i.e., $C^k_{\mathrm{b2s},1}(N_{\mathrm{sc}})=C^k_{\mathrm{s2u},1}(N_{\mathrm{sc}})$, and $C^k_{\mathrm{u2s},1}(N_{\mathrm{sc}})=C^k_{\mathrm{s2b},1}(N_{\mathrm{sc}})$. As a result, the optimal time allocation for the $k$th UE is
\begin{equation}
\mathrm{Downlink:} \left\lbrace 
\begin{array}{c}
T^k_{\mathrm{b2s}}=\frac{c^k_{\mathrm{s2u},1}\left( N_{\mathrm{sc}} \right)}{c^k_{\mathrm{b2s},1}\left( N_{\mathrm{sc}} \right) + c^k_{\mathrm{s2u},1}\left( N_{\mathrm{sc}} \right)}T_{\mathrm{dl}}, \\
T^k_{\mathrm{s2u}}=\frac{c^k_{\mathrm{b2s},1}\left( N_{\mathrm{sc}} \right)}{c^k_{\mathrm{b2s},1}\left( N_{\mathrm{sc}} \right) + c^k_{\mathrm{s2u},1}\left( N_{\mathrm{sc}} \right)}T_{\mathrm{dl}},
\end{array}
\right. 
\label{eq:time_dl_single}
\end{equation}
\begin{equation}
\mathrm{Uplink:} \left\lbrace 
\begin{array}{c}
T^k_{\mathrm{s2b}}=\frac{c^k_{\mathrm{u2s},1}\left( N_{\mathrm{sc}} \right)}{c^k_{\mathrm{s2b},1}\left( N_{\mathrm{sc}} \right)+c^k_{\mathrm{u2s},1}\left( N_{\mathrm{sc}} \right)}T_{\mathrm{ul}}, \\
T^k_{\mathrm{u2s}}=\frac{c^k_{\mathrm{s2b},1}\left( N_{\mathrm{sc}} \right)}{c^k_{\mathrm{s2b},1}\left( N_{\mathrm{sc}} \right)+c^k_{\mathrm{u2s},1}\left( N_{\mathrm{sc}} \right)}T_{\mathrm{ul}}.
\end{array}
\right. 
\label{eq:eq:time_ul_single}
\end{equation}
Unfortunately, (\ref{eq:time_dl_single}) and (\ref{eq:eq:time_ul_single}) are generally not hold for all BS-UE links unless the relations $c^k_{\mathrm{s2u},1}(N_{\mathrm{sc}})=c^k_{\mathrm{b2s},1}(N_{\mathrm{sc}})$ and $c^k_{\mathrm{u2s},1}(N_{\mathrm{sc}})=c^k_{\mathrm{s2b},1}(N_{\mathrm{sc}})$ are satisfied for all values of $k$, which is obviously not the case in practice. Hence, (\ref{eq:time_dl_single}) and (\ref{eq:eq:time_ul_single}) are only locally optimal for each BS-UE link. However, they could serve as the candidates of global optimization to maximize (\ref{eq:capacity_dl_b2u_ctdd}) and (\ref{eq:capacity_ul_u2b_ctdd}) respectively by exhaustive search of the $K$ candidates, denoted by CTDD-EXH. Alternatively, a suboptimal global selection, denoted by CTDD-SUB, which treats all BS-SC links as one backhaul link and all SC-UE links as one SC link, is 
\begin{equation}
\mathrm{Downlink:} \left\lbrace 
\begin{array}{c}
T_{\mathrm{b2s}}=\frac{\sum_{k=1}^K c^k_{\mathrm{s2u},1}\left( N_{\mathrm{sc}} \right)}{\sum_{k=1}^K c^k_{\mathrm{b2s},1}\left( N_{\mathrm{sc}} \right) + \sum_{k=1}^K c^k_{\mathrm{s2u},1}\left( N_{\mathrm{sc}} \right)}T_{\mathrm{dl}}, \\
T_{\mathrm{s2u}}=\frac{\sum_{k=1}^K c^k_{\mathrm{b2s},1}\left( N_{\mathrm{sc}} \right)}{\sum_{k=1}^K c^k_{\mathrm{b2s},1}\left( N_{\mathrm{sc}} \right) + \sum_{k=1}^K c^k_{\mathrm{s2u},1}\left( N_{\mathrm{sc}} \right)}T_{\mathrm{dl}},
\end{array}
\right. 
\label{eq:time_dl_subopt}
\end{equation}
\begin{equation}
\mathrm{Uplink:} \left\lbrace 
\begin{array}{c}
T_{\mathrm{s2b}}=\frac{\sum_{k=1}^K c^k_{\mathrm{u2s},1}\left( N_{\mathrm{sc}} \right)}{\sum_{k=1}^K c^k_{\mathrm{s2b},1}\left( N_{\mathrm{sc}} \right)+\sum_{k=1}^K c^k_{\mathrm{u2s},1}\left( N_{\mathrm{sc}} \right)}T_{\mathrm{ul}}, \\
T_{\mathrm{u2s}}=\frac{\sum_{k=1}^K c^k_{\mathrm{s2b},1}\left( N_{\mathrm{sc}} \right)}{\sum_{k=1}^K c^k_{\mathrm{s2b},1}\left( N_{\mathrm{sc}} \right)+\sum_{k=1}^K c^k_{\mathrm{u2s},1}\left( N_{\mathrm{sc}} \right)}T_{\mathrm{ul}}.
\end{array}
\right. 
\label{eq:eq:time_ul_subopt}
\end{equation}

\subsection{ZDD} \label{subsec:strategies_zdd}
If SCs are capable of ZDD, i.e., self-interference cancellation, they can transmit and receive signals in the same time-frequency resource. Under this assumption, since the communication between the BS and SCs dose not need to be separated from the data exchange between SCs and their related UEs, the overall throughput could be higher than the first strategy presented in Section \ref{subsec:strategies_ctdd}. In the following two subsections, perfect ZDD is assumed so that no Residual Self-Interference (RSI) exists. Assume that $N_{\mathrm{sct}}$ and $N_{\mathrm{scr}}$ antennas are used for transmitting and receiving in ZDD for each SC. Note that since the total number of antennas of each SC is $N_{\mathrm{sc}}$, then $N_{\mathrm{sct}} \leq N_{\mathrm{sc}}$ and $N_{\mathrm{scr}} \leq N_{\mathrm{sc}}$. 

For the downlink, the communication from the BS to each SC is similar to CTDD, only with $N_{\mathrm{scr}} \leq N_{\mathrm{sc}}$ receiving antennas at each SC. Then, for the $k$th SC, the average achievable throughput in $\mathrm{bit/s}$ is given by (\ref{eq:capacity_dl_b2s_wt}) with $N_{\mathrm{sc}}=N_{\mathrm{scr}}$, i.e., 
\begin{equation}
c^k_{\mathrm{b2s},2} \left( N_{\mathrm{scr}} \right) = c^k_{\mathrm{b2s},1}\left( N_{\mathrm{scr}} \right).
\label{eq:capacity_dl_b2s_zdd_wt}
\end{equation}
On the other hand, however, the data transmission from each SC to its related UE is now interfered by the corresponding BS-SC link. As a result, for the $k$th SC, its average achievable throughput is lower that (\ref{eq:capacity_dl_s2u_wt}) of CTDD even with $N_{\mathrm{sct}} = N_{\mathrm{sc}}$. Specifically, the term $\gamma_{\mathrm{s2u},s}(k)$ in (\ref{eq:wf_gamma}) changes to 
\begin{equation}
\gamma'_{\mathrm{s2u},s}\left( k \right) = \frac{a_{\mathrm{s2u},k} \lambda^2_s P_{\mathrm{sc}}}{B\sigma_{\mathrm{ue}}^2+I_{\mathrm{b2u}}\left( k \right)},
\label{eq:wf_s2u_gamma}
\end{equation}
where the interference power $I_{\mathrm{b2u}}(k)$ is 
\ifCLASSOPTIONonecolumn
\begin{equation}
I_{\mathrm{b2u}} \left( k \right)= \frac{P_{\mathrm{bs}} a_{\mathrm{b2u},k}}{N_{\mathrm{bs}}} \left(\max \{ \|\mathcal{H}^{\dagger}_{\mathrm{b2s},i} \| \}_{i=1}^{N_{\mathrm{bs}}} \right)^{-2} \| \mathbf{H}_{\mathrm{b2u},k} \mathcal{H}^{\dagger}_{\mathrm{b2s}} \|_\mathrm{F}^2  
\label{eq:itf_b2u_zdd}
\end{equation}
\else
\begin{align}
I_{\mathrm{b2u}} \left( k \right) & = \frac{P_{\mathrm{bs}} a_{\mathrm{b2u},k}}{N_{\mathrm{bs}}} \left(\max \{ \|\mathcal{H}^{\dagger}_{\mathrm{b2s},i} \| \}_{i=1}^{N_{\mathrm{bs}}} \right)^{-2} \nonumber \\ & \,\,\, \times \| \mathbf{H}_{\mathrm{b2u},k} \mathcal{H}^{\dagger}_{\mathrm{b2s}} \|_\mathrm{F}^2  
\label{eq:itf_b2u_zdd}
\end{align}
\fi
with $\| \cdot \|_\mathrm{F}$ denoting the Frobenius norm. Hence, the water-filling power allocation of (\ref{eq:wf_s2u}) changes to    
\begin{equation}
\renewcommand{\arraycolsep}{2pt} 
\frac{P'_{\mathrm{sc},s}\left( k \right)}{P_{\mathrm{sc}}} = \left\lbrace 
\begin{array}{cc}
\frac{1}{\gamma'_{\mathrm{s2u},0}\left( k \right)} - \frac{1}{\gamma'_{\mathrm{s2u},s}\left( k \right)}, & \gamma'_{\mathrm{s2u},s}\left( k \right) \geq \gamma'_{\mathrm{s2u},0}\left( k \right), \\
0, & \gamma'_{\mathrm{s2u},s}\left( k \right) < \gamma'_{\mathrm{s2u},0}\left( k \right),
\end{array}
\right.
\label{eq:wf_s2u_zdd}
\end{equation}
for the communication of $k$th SC-UE link. Assume that $S'_{\mathrm{s2u}}(k) \in \{1,\ldots,\min\{N_{\mathrm{sct}},N_{\mathrm{ue}}\}\}$ is the number of parallel streams supported by the water-filling power allocation (\ref{eq:wf_s2u_zdd}) \cite{Goldsmith_WC_Ch10}. Then, from the $k$th SC to its associated UE, the average achievable throughput in $\mathrm{bit/s}$ is    
\begin{equation}
c^k_{\mathrm{s2u},2} \left( N_{\mathrm{sct}} \right) = \mathsf{E} \left\lbrace B \sum^{S'_{\mathrm{s2u}}\left( k \right)}_{s=1} \log_2\left[ \frac{\gamma'_{\mathrm{s2u},s}\left( k \right)}{\gamma'_{\mathrm{s2u},0}\left( k \right)} \right] \right\rbrace.
\label{eq:capacity_dl_s2u_zdd}
\end{equation}

As for the uplink, the data transmission from UEs to their related SCs is similar to CTDD, only with $N_{\mathrm{scr}} \leq N_{\mathrm{sc}}$. As a result, for the $k$th SC, the average achievable throughput is given by (\ref{eq:capacity_ul_u2s_wt}) with $N_{\mathrm{sc}}=N_{\mathrm{scr}}$ as 
\begin{equation}
c^k_{\mathrm{u2s},2} \left(N_{\mathrm{scr}}\right) = c^k_{\mathrm{u2s},1}\left(N_{\mathrm{scr}}\right).
\label{eq:capacity_ul_u2s_zdd_wt}
\end{equation}
On the other hand, the communication from SCs to the BS is now interfered by all UE-SC links. Therefore, for the $k$th SC, its average achievable throughput is lower than (\ref{eq:capacity_ul_s2b_wt}) of CTDD even with $N_{\mathrm{sct}} = N_{\mathrm{sc}}$. Specifically, the interference power for the $i$th stream of the $k$th SC $I_{\mathrm{u2b}}( i, k )$ is 
\begin{equation}
I_{\mathrm{u2b}} \left( i, k \right) = P_{\mathrm{ue}} \|(\mathcal{H}_{\mathrm{b2u}} \mathcal{H}^{\dagger}_{\mathrm{b2s}})^\mathrm{T}_j \|^{2} 
\label{eq:itf_u2b_zdd}
\end{equation}
where $j=i+(k-1)N_{\mathrm{sct}}$ with $i=1,\ldots, N_{\mathrm{sct}}$. Then, from the $k$th SC to the BS, the average achievable throughput is   
\ifCLASSOPTIONonecolumn
\begin{equation}
c^k_{\mathrm{s2b},2} \left(N_{\mathrm{sct}}\right) = \mathsf{E} \left\lbrace B \sum_{i=1}^{N_{\mathrm{sct}}}\log_2\left( \frac{P_{\mathrm{sc}}}{N_{\mathrm{sct}}} \left[B\sigma^2_{\mathrm{bs}} \|(\mathcal{H}^{\dagger}_{\mathrm{b2s}})^\mathrm{T}_j \|^{2} + I_{\mathrm{u2b}} \left( i, k \right) \right]^{-1}+1 \right) \right\rbrace.
\label{eq:capacity_ul_s2b_zdd}
\end{equation}
\else
\begin{align}
c^k_{\mathrm{s2b},2} \left(N_{\mathrm{sct}}\right) & = \mathsf{E} \left\lbrace B \sum_{i=1}^{N_{\mathrm{sct}}}\log_2\left( \frac{P_{\mathrm{sc}}}{N_{\mathrm{sct}}} \left[B\sigma^2_{\mathrm{bs}} \|(\mathcal{H}^{\dagger}_{\mathrm{b2s}})^\mathrm{T}_j \|^{2} \right. \right. \right. \nonumber \\ & \left. \left. \left. \,\,\, + I_{\mathrm{u2b}} \left( i, k \right) \right]^{-1}+1 \right) \right\rbrace.
\label{eq:capacity_ul_s2b_zdd}
\end{align}
\fi

For ZDD, both the downlink throughput and uplink throughput are limited by the weaker one of their related two communication parts. Therefore, the downlink average achievable throughput of ZDD removing resources not used for transmission for the $k$th UE is
\ifCLASSOPTIONonecolumn
\begin{equation}
C^{\mathrm{dl},k}_{2}\left( N_{\mathrm{sct}}, N_{\mathrm{scr}} \right) = \left(\frac{T_{\mathrm{dl}}}{T}\right) \min\left\lbrace c^k_{\mathrm{b2s},1}\left( N_{\mathrm{scr}} \right), c^k_{\mathrm{s2u},2}\left( N_{\mathrm{sct}} \right) \right\rbrace.
\label{eq:capacity_dl_b2u_zdd}
\end{equation}
\else
\begin{align}
C^{\mathrm{dl},k}_{2}\left( N_{\mathrm{sct}}, N_{\mathrm{scr}} \right) & = \left(\frac{T_{\mathrm{dl}}}{T}\right) \min\left\lbrace c^k_{\mathrm{b2s},1}\left( N_{\mathrm{scr}} \right), \right . \nonumber \\ & \,\,\,\,\,\, \left. c^k_{\mathrm{s2u},2}\left( N_{\mathrm{sct}} \right) \right\rbrace.
\label{eq:capacity_dl_b2u_zdd}
\end{align}
\fi
On the other hand, the uplink average achievable throughput of ZDD removing resources not used for transmission for the $k$th UE can be written as
\ifCLASSOPTIONonecolumn
\begin{equation}
C^{\mathrm{ul},k}_{2}\left( N_{\mathrm{sct}}, N_{\mathrm{scr}} \right) = \left(\frac{T_{\mathrm{ul}}}{T}\right) \min\left\lbrace c^k_{\mathrm{u2s},1} \left( N_{\mathrm{scr}} \right),  c^k_{\mathrm{s2b},2} \left(N_{\mathrm{sct}}\right) \right\rbrace.
\label{eq:capacity_ul_b2u_zdd}
\end{equation}
\else
\begin{align}
C^{\mathrm{ul},k}_{2}\left( N_{\mathrm{sct}}, N_{\mathrm{scr}} \right) & = \left(\frac{T_{\mathrm{ul}}}{T}\right)  \min\left\lbrace c^k_{\mathrm{u2s},1} \left( N_{\mathrm{scr}} \right), \right. \nonumber \\ & \,\,\,\,\,\, \left. c^k_{\mathrm{s2b},2} \left(N_{\mathrm{sct}}\right) \right\rbrace.
\label{eq:capacity_ul_b2u_zdd}
\end{align}
\fi

\subsection{ZDD-IR} \label{subsec:strategies_zdd-ir}
As shown in Section \ref{subsec:strategies_zdd}, the interference in ZDD reduces $c^k_{\mathrm{s2u},1}(N_{\mathrm{sc}})$ and $c^k_{\mathrm{s2b},1}(N_{\mathrm{sc}})$ to $c^k_{\mathrm{s2u},2}(N_{\mathrm{sct}})$ and $c^k_{\mathrm{s2b},2}(N_{\mathrm{sct}})$ respectively, which would limit the overall throughput. Since $N_{\mathrm{bs}} \gg KN_{\mathrm{ue}}$, the interference can be rejected at the BS by a $N_{\mathrm{bs}} \times (N_{\mathrm{bs}} - KN_{\mathrm{ue}} )$ matrix $\mathbf{R}$ which satisfies  
\begin{equation}
\mathcal{H}_{\mathrm{b2u}} \mathbf{R} = \mathbf{0}_{KN_{\mathrm{ue}} \times \left(N_{\mathrm{bs}} - KN_{\mathrm{ue}} \right)},
\label{eq:ir} 
\end{equation}
where $\mathbf{0}$ denotes the all-zero matrix.

In the case of downlink, the ZF-IR beamforming matrix is 
\begin{equation}
\mathbf{G} = \mathbf{R} \left(\mathcal{H}_{\mathrm{b2s}}\mathbf{R}\right)^{\dagger}.
\label{eq:zf-ir matrix} 
\end{equation}
Note that $(\mathcal{H}_{\mathrm{b2s}}\mathbf{R})^{\dagger}$ is a valid right inverse when $N_{\mathrm{bs}} \geq K(N_{\mathrm{ue}}+N_{\mathrm{scr}})$. Note that in massive MIMO systems, because the assumption $N_{\mathrm{bs}} \gg KN_{\mathrm{sc}}$ mentioned in Section \ref{subsec:strategies_ctdd}, the fact $N_{\mathrm{scr}} \le N_{\mathrm{sc}}$ mentioned in Section \ref{subsec:strategies_zdd}, and the common practical assumption $N_{\mathrm{sc}} \ge N_{\mathrm{ue}}$, the assumption $N_{\mathrm{bs}} \geq K(N_{\mathrm{ue}}+N_{\mathrm{scr}})$ is highly probable in practice. Nevertheless, $N_{\mathrm{bs}} \geq K(N_{\mathrm{ue}}+N_{\mathrm{scr}})$ is a requirement for ZDD-IR. For the communication from the $k$th SC to its related UE, the interference power in (\ref{eq:itf_b2u_zdd}) becomes  
\begin{equation}
I'_{\mathrm{b2u}} \left( k \right) = \frac{P_{\mathrm{bs}} a_{\mathrm{b2u},k}}{N_{\mathrm{bs}}} \left(\max \{ \|\mathcal{H}^{\dagger}_{\mathrm{b2s},i} \| \}_{i=1}^{N_{\mathrm{bs}}} \right)^{-2} \| \mathbf{H}_{\mathrm{b2u},k} \mathbf{G} \|_\mathrm{F}^2.  
\label{eq:itf_b2u_zdd-ir}
\end{equation}
Note that 
\ifCLASSOPTIONonecolumn
\begin{equation}
\mathbf{H}_{\mathrm{b2u},k} \mathbf{G} = \left( \mathbf{H}_{\mathrm{b2u},k}  \mathbf{R} \right) \left(\mathcal{H}_{\mathrm{b2s}}\mathbf{R}\right)^{\dagger} = \mathbf{0}_{N_{\mathrm{ue}}\times \left(N_{\mathrm{bs}} - KN_{\mathrm{ue}} \right)} \left(\mathcal{H}_{\mathrm{b2s}}\mathbf{R}\right)^{\dagger} = \mathbf{0}_{N_{\mathrm{ue}} \times KN_{\mathrm{scr}}}.
\label{eq:ir_b2u}
\end{equation}
\else
\begin{align}
\mathbf{H}_{\mathrm{b2u},k} \mathbf{G} & = \left( \mathbf{H}_{\mathrm{b2u},k}  \mathbf{R} \right) \left(\mathcal{H}_{\mathrm{b2s}}\mathbf{R}\right)^{\dagger} \nonumber \\ & = \mathbf{0}_{N_{\mathrm{ue}}\times \left(N_{\mathrm{bs}} - KN_{\mathrm{ue}} \right)} \left(\mathcal{H}_{\mathrm{b2s}}\mathbf{R}\right)^{\dagger} \nonumber \\ & = \mathbf{0}_{N_{\mathrm{ue}} \times KN_{\mathrm{scr}}}.
\label{eq:ir_b2u}
\end{align}
\fi
Based on (\ref{eq:ir_b2u}), the interference power in (\ref{eq:itf_b2u_zdd-ir}) is then zero. As a result, from the $k$th SC to its associated UE, the average achievable throughput in $\mathrm{bit/s}$ is similar to CTDD only with $N_{\mathrm{sct}} \leq N_{\mathrm{sc}}$, which is given by (\ref{eq:capacity_dl_s2u_wt}) with $N_{\mathrm{sc}}=N_{\mathrm{sct}}$ as 
\begin{equation}
c^k_{\mathrm{s2u},3} \left( N_{\mathrm{sct}} \right) = c^k_{\mathrm{s2u},1} \left( N_{\mathrm{sct}} \right).
\label{eq:capacity_dl_s2u_zdd-ir_wt}
\end{equation}
On the other hand, from the BS to the $k$th SC, the downlink average achievable throughput in $\mathrm{bit/s}$ becomes
\ifCLASSOPTIONonecolumn
\begin{equation}
c^k_{\mathrm{b2s},3}\left( N_{\mathrm{scr}} \right) = \mathsf{E} \left\lbrace B N_{\mathrm{scr}} \log_2\left[ \frac{P_{\mathrm{bs}}}{BN_{\mathrm{bs}}\sigma^2_{\mathrm{sc}} } \left(\max \{ \|\mathbf{g}_i \| \}_{i=1}^{N_{\mathrm{bs}}} \right)^{-2} +1 \right] \right\rbrace,
\label{eq:capacity_dl_b2s_zdd-ir_wt}
\end{equation}
\else
\begin{align}
c^k_{\mathrm{b2s},3}\left( N_{\mathrm{scr}} \right) & = \mathsf{E} \left\lbrace B N_{\mathrm{scr}} \log_2\left[ \frac{P_{\mathrm{bs}}}{BN_{\mathrm{bs}}\sigma^2_{\mathrm{sc}} } \right. \right. \nonumber \\ & \left. \left. \,\,\, \times \left(\max \{ \|\mathbf{g}_i \| \}_{i=1}^{N_{\mathrm{bs}}} \right)^{-2} +1 \right] \right\rbrace,
\label{eq:capacity_dl_b2s_zdd-ir_wt}
\end{align}
\fi
where $\mathbf{g}_i$ is the $i$th row of $\mathbf{G}$. 

As for the uplink, the ZF-IR decoding matrix is $\mathbf{G}^\mathrm{T}$ with the requirement $N_{\mathrm{bs}} \geq K(N_{\mathrm{ue}}+N_{\mathrm{sct}})$. Similarly to the downlink, the assumption $N_{\mathrm{bs}} \geq K(N_{\mathrm{ue}}+N_{\mathrm{sct}})$ is highly probable in practice. For the data transmission from the $k$th UE to its associated SC, the average achievable throughput in $\mathrm{bit/s}$ is still the same as in ZDD, i.e.,
\begin{equation}
c^k_{\mathrm{u2s},3}\left(N_{\mathrm{scr}}\right) = c^k_{\mathrm{u2s},2} \left( N_{\mathrm{scr}} \right) = c^k_{\mathrm{u2s},1} \left( N_{\mathrm{scr}} \right).
\label{eq:capacity_ul_u2s_zdd-ir_wt}
\end{equation}
On the other hand, in the case of the communication from the $k$th SC to the BS, the interference power for the $i$th stream of the $k$th SC in (\ref{eq:itf_u2b_zdd}) becomes 
\begin{equation}
I'_{\mathrm{u2b}} \left( i, k \right) = P_{\mathrm{ue}} \|(\mathcal{H}_{\mathrm{b2u}} \mathbf{G})^\mathrm{T}_j \|^{2} 
\label{eq:itf_u2b_zdd-ir}
\end{equation}
where $j=i+(k-1)N_{\mathrm{sct}}$ with $i=1,\ldots, N_{\mathrm{sct}}$. Similarly to (\ref{eq:ir_b2u}) of the downlink case,
\ifCLASSOPTIONonecolumn
\begin{equation}
\mathcal{H}_{\mathrm{b2u}} \mathbf{G} = \mathbf{A}_{\mathrm{b2u}} \left( \mathbf{H}_{\mathrm{b2u}}  \mathbf{R} \right) \left(\mathcal{H}_{\mathrm{b2s}}\mathbf{R}\right)^{\dagger} =\mathbf{A}_{\mathrm{b2u}} \mathbf{0}_{KN_{\mathrm{ue}}\times \left(N_{\mathrm{bs}} - KN_{\mathrm{ue}} \right)} \left(\mathcal{H}_{\mathrm{b2s}}\mathbf{R}\right)^{\dagger} = \mathbf{0}_{KN_{\mathrm{ue}} \times KN_{\mathrm{sct}}}.
\label{eq:ir_u2b}
\end{equation}
\else
\begin{align}
\mathcal{H}_{\mathrm{b2u}} \mathbf{G} & = \mathbf{A}_{\mathrm{b2u}} \left( \mathbf{H}_{\mathrm{b2u}}  \mathbf{R} \right) \left(\mathcal{H}_{\mathrm{b2s}}\mathbf{R}\right)^{\dagger} \nonumber \\ & =\mathbf{A}_{\mathrm{b2u}} \mathbf{0}_{KN_{\mathrm{ue}}\times \left(N_{\mathrm{bs}} - KN_{\mathrm{ue}} \right)} \left(\mathcal{H}_{\mathrm{b2s}}\mathbf{R}\right)^{\dagger} \nonumber \\ & = \mathbf{0}_{KN_{\mathrm{ue}} \times KN_{\mathrm{sct}}}.
\label{eq:ir_u2b}
\end{align}
\fi
According to (\ref{eq:ir_u2b}), the interference power in (\ref{eq:itf_u2b_zdd-ir}) is then zero. Therefore, from the $k$th SC to the BS, the average achievable throughput in $\mathrm{bit/s}$ becomes 
\ifCLASSOPTIONonecolumn
\begin{equation}
c^k_{\mathrm{s2b},3} \left(N_{\mathrm{sct}}\right) = \mathsf{E} \left\lbrace B \sum_{i=1}^{N_{\mathrm{sct}}}\log_2\left[ \frac{P_{\mathrm{sc}}}{N_{\mathrm{sct}}} \left(B\sigma^2_{\mathrm{bs}} \|\mathbf{g}^\mathrm{T}_j \|^{2} \right)^{-1}+1 \right] \right\rbrace,
\label{eq:capacity_ul_s2b_zdd-ir_wt}
\end{equation}
\else
\begin{align}
c^k_{\mathrm{s2b},3} \left(N_{\mathrm{sct}}\right) & = \mathsf{E} \left\lbrace B \sum_{i=1}^{N_{\mathrm{sct}}}\log_2\left[ \frac{P_{\mathrm{sc}}}{N_{\mathrm{sct}}} \right. \right. \nonumber \\ & \left. \left. \,\,\, \times \left(B\sigma^2_{\mathrm{bs}} \|\mathbf{g}^\mathrm{T}_j \|^{2} \right)^{-1}+1 \right] \right\rbrace,
\label{eq:capacity_ul_s2b_zdd-ir_wt}
\end{align}
\fi
where $\mathbf{g}^T_i$ is the $i$th column of $\mathbf{G}$.

According to the discussion above, the downlink average achievable throughput of ZDD-IR removing resources not used for transmission for the $k$th UE is
\ifCLASSOPTIONonecolumn
\begin{equation}
C^{\mathrm{dl},k}_{3}\left( N_{\mathrm{sct}}, N_{\mathrm{scr}} \right) = \left(\frac{T_{\mathrm{dl}}}{T} \right) \min\left\lbrace c^k_{\mathrm{b2s},3}\left(N_{\mathrm{scr}}\right), c^k_{\mathrm{s2u},1}\left( N_{\mathrm{sct}} \right) \right\rbrace.
\label{eq:capacity_dl_b2u_zdd-ir}
\end{equation}
\else
\begin{align}
C^{\mathrm{dl},k}_{3}\left( N_{\mathrm{sct}}, N_{\mathrm{scr}} \right) & = \left(\frac{T_{\mathrm{dl}}}{T} \right) \min\left\lbrace c^k_{\mathrm{b2s},3}\left(N_{\mathrm{scr}}\right), \right. \nonumber \\\ 
& \,\,\,\,\,\, \left. c^k_{\mathrm{s2u},1}\left( N_{\mathrm{sct}} \right) \right\rbrace.
\label{eq:capacity_dl_b2u_zdd-ir}
\end{align}
\fi
On the other hand, the uplink average achievable throughput of ZDD-IR removing resources not used for transmission for the $k$th UE can be written as
\ifCLASSOPTIONonecolumn
\begin{equation}
C^{\mathrm{ul},k}_{3}\left( N_{\mathrm{sct}}, N_{\mathrm{scr}} \right) = \left(\frac{T_{\mathrm{ul}}}{T} \right) \min\left\lbrace c^k_{\mathrm{u2s},1} \left( N_{\mathrm{scr}} \right),  c^k_{\mathrm{s2b},3} \left(N_{\mathrm{sct}}\right) \right\rbrace.
\label{eq:capacity_ul_b2u_zdd-ir}
\end{equation}
\else
\begin{align}
C^{\mathrm{ul},k}_{3}\left( N_{\mathrm{sct}}, N_{\mathrm{scr}} \right) & = \left(\frac{T_{\mathrm{ul}}}{T} \right) \min\left\lbrace c^k_{\mathrm{u2s},1} \left( N_{\mathrm{scr}} \right), \right. \nonumber \\ & \,\,\,\,\,\, \left. c^k_{\mathrm{s2b},3} \left(N_{\mathrm{sct}}\right) \right\rbrace.
\label{eq:capacity_ul_b2u_zdd-ir}
\end{align}
\fi
\section{Simulation Results} \label{sec:results}

\begin{figure}[!t]
\centering \includegraphics[width = 1.0\linewidth]{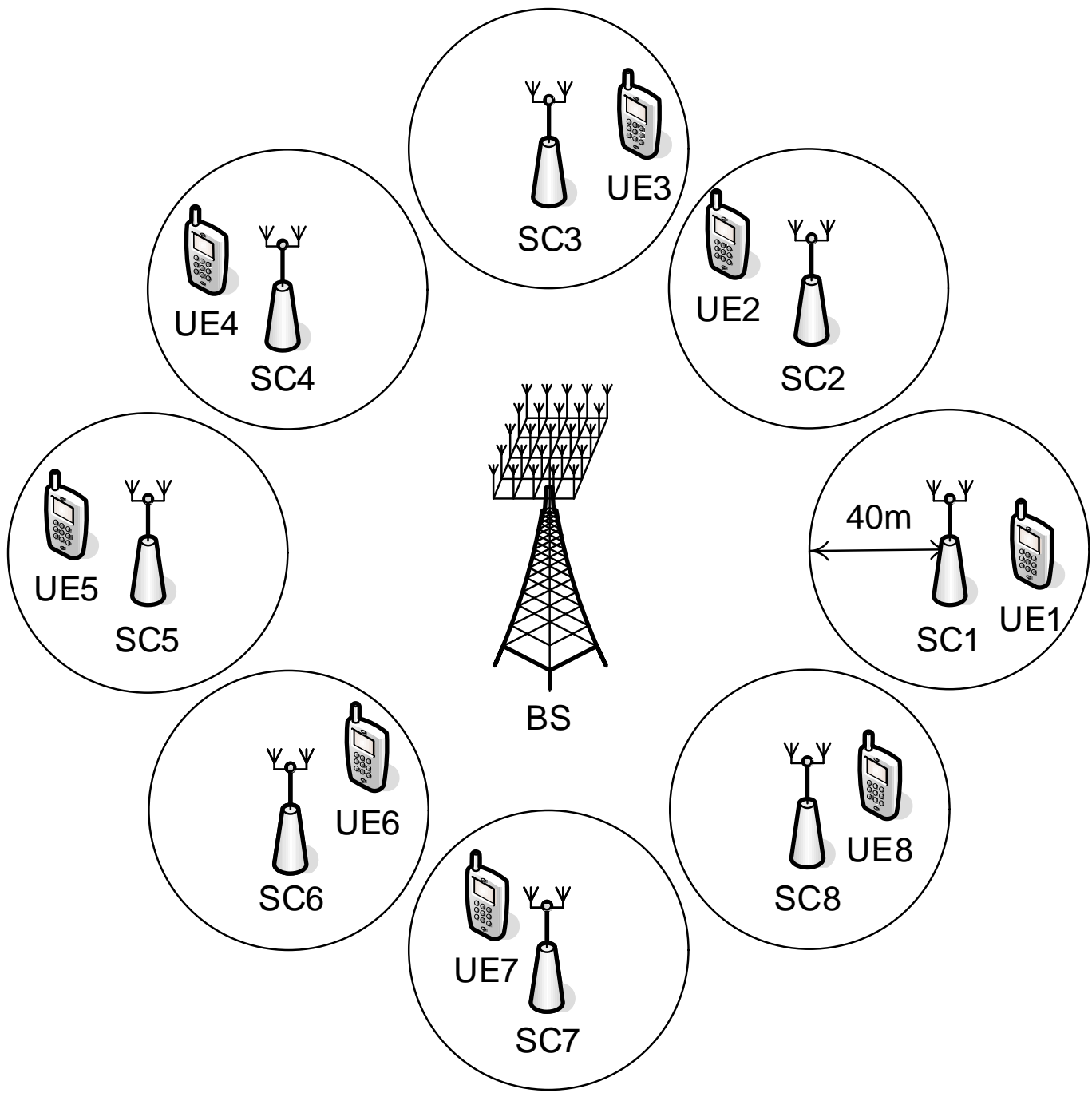}
\caption{System diagram for simulations.}
\label{fig:system}
\end{figure}

Fig. \ref{fig:system} presents the diagram of the considered massive MIMO system for simulations. Specifically, a BS supports $K=8$ UEs, each of which is associated to a SC. The distances between the BS and SCs are considered to be the same. SCs are evenly distributed, and each of them covers a disc area with the radius of $40$ meters as in \cite{3GPP_TR_36.828}. The closest distance between a UE to its related SC is $10$ meters also as in \cite{3GPP_TR_36.828}, and each UE is uniformly distributed within the service area of its associated SC. In Table \ref{table:sim_parameters}, simulation parameters of the considered massive MIMO system are provided. To exploit the advantages of multiple antennas, the Non-Line-of-Sight (NLoS) models with shadowing at the carrier frequency $f_{\mathrm{c}}=2\mathrm{GHz}$ provided in \cite{3GPP_TR_36.828} are employed. The maximal power and noise power are based on \cite{3GPP_TR_36.828} as well. The bandwidth $B=20\mathrm{MHz}$ is a particular number in LTE/LTE-A \cite{3GPP_TS_36.201}. The noise figure values are based on \cite{R4-092042,3GPP_TR_36.828}, and the antenna gains for SCs and UEs are based on \cite{3GPP_TR_36.828}. The BS antenna gain is considered as $2\mathrm{dBi}$ due to the small antennas supposed to be applied in massive MIMO systems. The downlink and uplink are separately operated in TDD mode. Equal time fractions of downlink and uplink transmissions are assumed and the overhead time is not considered in simulations, i.e., $T_{\mathrm{dl}}/T=T_{\mathrm{ul}}/T=0.5$. For each scenario, SC in-band wireless backhaul systems employing CTDD-EXH, CTDD-SUB, ZDD, and ZDD-IR techniques discussed in Section \ref{sec:strategies} are considered, and their sum-rate values are calculated. In the result figures, $N_{\mathrm{sc}}=\{N_{\mathrm{sct}},N_{\mathrm{scr}}\}$ means that $N_{\mathrm{sct}}$ and $N_{\mathrm{scr}}$ antennas are employed to transmit and receive signals in the same time-frequency resource for ZDD and ZDD-IR. Note that $N_{\mathrm{sc}} \geq N_{\mathrm{sct}} \geq N_{\mathrm{ue}}$ and $N_{\mathrm{sc}} \geq N_{\mathrm{scr}} \geq N_{\mathrm{ue}}$ are assumed.

\ifCLASSOPTIONonecolumn
\begin{table}[!t]
\renewcommand{\arraystretch}{1.3}
\caption{Simulation Parameters}
\centering
\begin{tabular}{|c|c|}
\hline
Carrier Frequency & $f_{\mathrm{c}}=2\mathrm{GHz}$ \\
\hline
System Bandwidth & $B=20\mathrm{MHz}$ \\
\hline
Number of SCs & $K=8$ \\
\hline
Cell Type & BS: Macro; SC: Outdoor Pico \\
\hline
Channel Model & Large-Scale: NLoS Models in \cite{3GPP_TR_36.828}; Small-Scale: Rayleigh Fading \\
\hline
Maximal Power & $P_{\mathrm{bs}}=35\mathrm{w \, (45.4dBm)}$; $P_{\mathrm{sc}}=250\mathrm{mw \,(24dBm)}$; $P_{\mathrm{ue}}=200\mathrm{mw \, (23dBm)}$ \\
\hline
Antenna Gain & BS: $2\mathrm{dBi}$; SC: $5\mathrm{dBi}$; UE: $0\mathrm{dBi}$ \\
\hline
Number of Antennas & $N_{\mathrm{bs}}=256$; $N_{\mathrm{sc}}=4$; $N_{\mathrm{ue}} \in \{1,2\}$ \\
\hline
Noise Power & $N_0=-174\mathrm{dBm/Hz}$ \\
\hline
Noise Figure & BS: $5\mathrm{dB}$; SC: $8\mathrm{dB}$; UE: $9\mathrm{dB}$ \\
\hline
\end{tabular}
\label{table:sim_parameters}
\end{table}
\else
\vspace*{4pt}
\begin{table*}[!t]
\normalsize
\hrulefill
\renewcommand{\arraystretch}{1.3}
\caption{Simulation Parameters}
\centering
\begin{tabular}{|c|c|}
\hline
Carrier Frequency & $f_{\mathrm{c}}=2\mathrm{GHz}$ \\
\hline
System Bandwidth & $B=20\mathrm{MHz}$ \\
\hline
Number of SCs& $K=8$ \\
\hline
Cell Type & BS: Macro; SC: Outdoor Pico \\
\hline
Channel Model & Large-Scale: NLoS Models in \cite{3GPP_TR_36.828}; Small-Scale: Rayleigh Fading \\
\hline
Maximal Power & $P_{\mathrm{bs}}=35\mathrm{w \, (45.4dBm)}$; $P_{\mathrm{sc}}=250\mathrm{mw \,(24dBm)}$; $P_{\mathrm{ue}}=200\mathrm{mw \, (23dBm)}$ \\
\hline
Antenna Gain & BS: $2\mathrm{dBi}$; SC: $5\mathrm{dBi}$; UE: $0\mathrm{dBi}$ \\
\hline
Number of Antennas & $N_{\mathrm{bs}}=256$; $N_{\mathrm{sc}}=4$; $N_{\mathrm{ue}} \in \{1,2\}$ \\
\hline
Noise Power & $N_0=-174\mathrm{dBm/Hz}$ \\
\hline
Noise Figure & BS: $5\mathrm{dB}$; SC: $8\mathrm{dB}$; UE: $9\mathrm{dB}$ \\
\hline
\end{tabular}
\label{table:sim_parameters}
\end{table*}
\fi

\subsection{Downlink} \label{subsec:results_dl}
\begin{figure}[!t]
\centering \includegraphics[width = 1.0\linewidth]{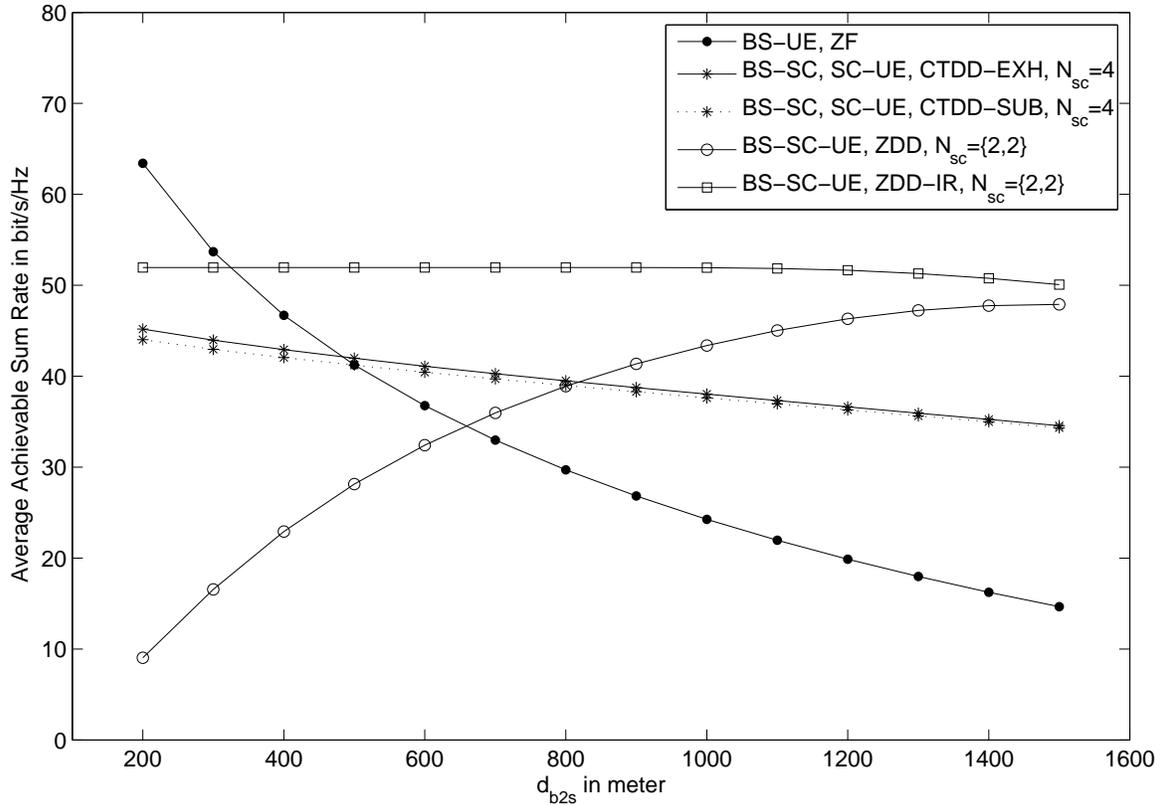}
\caption{Downlink average achievable sum rates of different techniques for $N_{\mathrm{sc}}=4$ and $N_{\mathrm{ue}}=1$ with different $d_{\mathrm{b2s}}$ values.}
\label{fig:dl_Nue1_Nsc4}
\end{figure}

\begin{figure}[!t]
\centering \includegraphics[width = 1.0\linewidth]{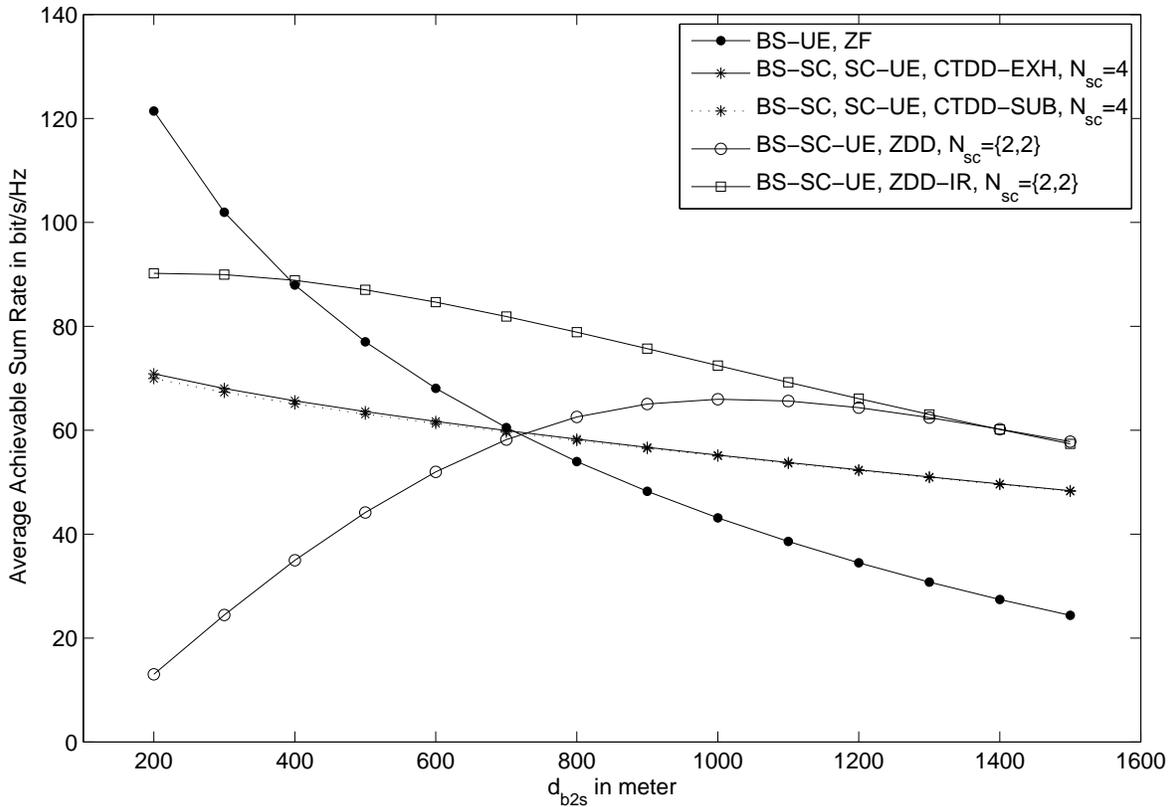}
\caption{Downlink average achievable sum rates of different techniques for $N_{\mathrm{sc}}=4$ and $N_{\mathrm{ue}}=2$ with different $d_{\mathrm{b2s}}$ values.}
\label{fig:dl_Nue2_Nsc4}
\end{figure}

In the case of downlink scenario, the distance between the BS and each SC is considered to be at the range of $d_{\mathrm{b2s}} \in\{200\mathrm{m}, 300\mathrm{m}, \ldots, 1500\mathrm{m}\}$. The average achievable downlink sum-rate results in $\mathrm{bit/s/Hz}$ employing different techniques discussed in Section \ref{sec:strategies} are compared to the results of directly ZF beamforming without SCs discussed in Section \ref{sec:basis}. Note that a larger $d_{\mathrm{b2s}}$ value implies worse BS-SC and BS-UE links. For $N_{\mathrm{ue}}=1$, Fig. \ref{fig:dl_Nue1_Nsc4} shows the average achievable sum-rate results of $N_{\mathrm{sc}}=4$. Similarly, as for $N_{\mathrm{ue}}=2$, the case of $N_{\mathrm{sc}}=4$ is presented in Fig. \ref{fig:dl_Nue2_Nsc4}.

Compared to directly ZF beamforming without SCs, although CTDD-EXH separates the downlink into two parts with shorter times, with the help of multiple antennas at SCs and the better links of BS-SC and SC-UE, it could achieve a significantly better average achievable sum rate, when the BS-UE links are not very strong as shown in the figures. In addition, it can be expected that a larger $N_{\mathrm{sc}}$ could offer a greater average achievable sum rate. Note that both techniques achieve a lower average achievable sum rate as the BS-SC links become worse, but directly ZF beamforming decays much quicker. As a result, if the BS-UE links are sufficiently strong, directly ZF beamforming could provide the better average achievable sum rate as shown in the figures. Nevertheless, by increasing $N_{\mathrm{sc}}$, it could be expected that CTDD-EXH always has the potential to outperform directly ZF beamforming, as long as the SC-UE links are better than the BS-UE links, which is mostly the practical case. Moreover, the suboptimal CTDD-SUB could provide almost the same results as CTDD-EXH.

For ZDD, as the BS-SC links become weaker, it offers the better average achievable sum rate up to a point, from where the average achievable sum rate starts to drop. This trend can be clearly read from Fig. \ref{fig:dl_Nue2_Nsc4}. The reason is that the BS-SC links interfere the SC-UE links since they are operated at the same time-frequency resource. When the interference is relatively strong, the SC-UE links limit the overall average achievable sum rate. The interference decreases as the BS-SC links become weaker, then the SC-UE links become better. Therefore, the average achievable sum rate becomes greater. However, when the SC-UE links are strong enough, the overall average achievable sum rate is then limited by the BS-SC links. As a result, the overall average achievable sum rate starts to decrease. Note that in Fig. \ref{fig:dl_Nue1_Nsc4} it has not reached the maximum point yet. Due to the fact that the BS-SC links are highly probably better than the BS-UE links if $d_{\mathrm{b2s}}$ is not too small, when the BS-SC links limit the average achievable sum rate, it is still much better than directly ZF beamforming, with the condition of $N_{\mathrm{scr}} \geq N_{\mathrm{ue}}$. As a result, after a certain value of $d_{\mathrm{b2s}}$, ZDD is better than directly ZF beamforming without SCs as shown in the figures. Note that in the case of complete ZDD, which means that each SC antenna can be used for transmitting and receiving signals in the same time-frequency resource, when the BS-SC links limit the average achievable sum rate, complete ZDD is always better than CTDD-EXH. However, in this subsection, the more conservative ZDD method in which separate transmitting and receiving antennas are considered. In this case, the comparison between ZDD and CTDD-EXH is more complicated. Nevertheless, the conservative ZDD still could outperform CTDD-EXH as shown in the figures.

In the case of ZDD-IR, as the interference of the SC-UE links from the BS-SC links is rejected, it is generally better than ZDD as shown in the figures. For the range that the average achievable sum rate is limited by the SC-UE links, it is independent of $d_{\mathrm{b2s}}$ as shown in Fig. \ref{fig:dl_Nue1_Nsc4}. On the other hand, for the range that the average achievable sum rate is limited by the BS-SC links, it decreases as $d_{\mathrm{b2s}}$ increases, as shown in the figures. Eventually, ZDD-IR becomes similar to ZDD with slightly worse average achievable sum rate as shown in Fig. \ref{fig:dl_Nue2_Nsc4}. The reason is that the interference rejection process could result in slightly worse BS-SC links due to the power constraint for each antenna.  

To sum up, if SCs are located very close to the BS, directly ZF beamforming between BS and UEs is the best choice, which means that employing SCs in this case is pointless. On the other hand, when SCs are located with certain distances to the BS, all CTDD, ZDD, and ZDD-IR could outperform directly ZF beamforming without SCs. Specifically, ZDD-IR is the best choice for the middle range, while ZDD is good enough for the far range.

\subsection{Uplink} \label{subsec:results_ul}
\begin{figure}[!t]
\centering \includegraphics[width = 1.0\linewidth]{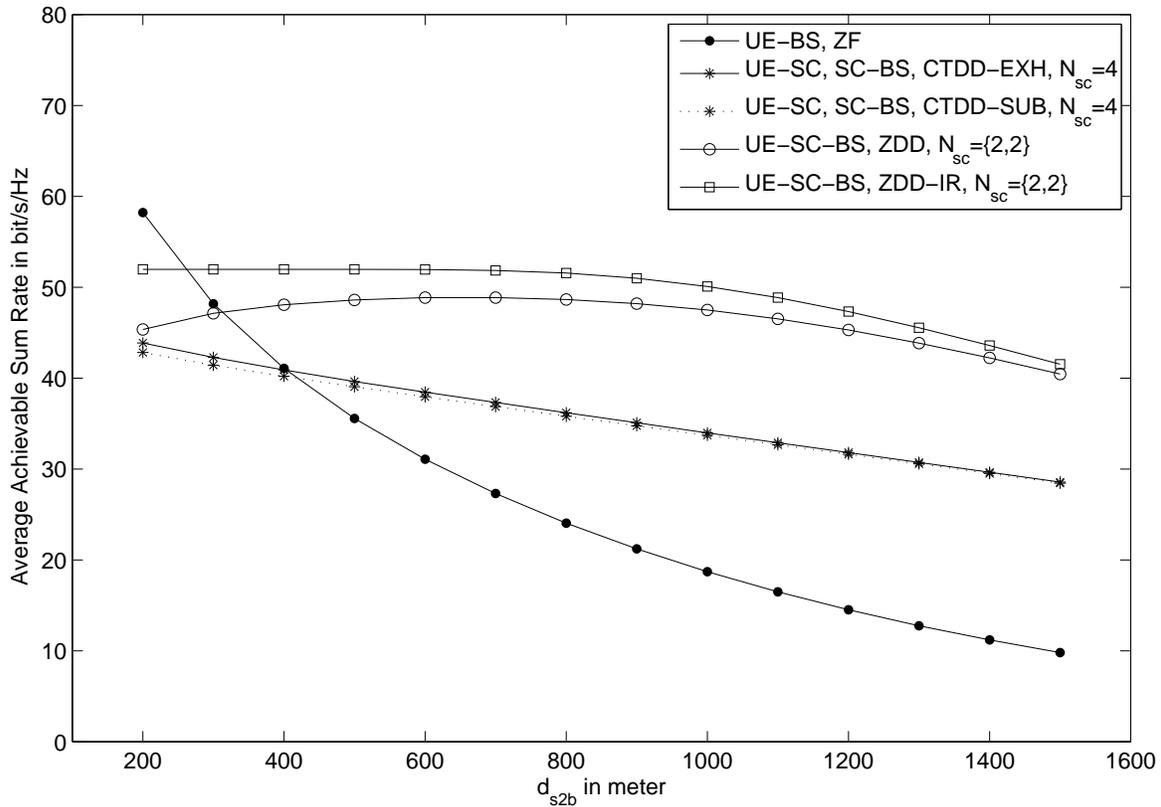}
\caption{Uplink average achievable sum rates of different techniques for $N_{\mathrm{sc}}=4$ and $N_{\mathrm{ue}}=1$ with different $d_{\mathrm{s2b}}$ values.}
\label{fig:ul_Nue1_Nsc4}
\end{figure}

\begin{figure}[!t]
\centering \includegraphics[width = 1.0\linewidth]{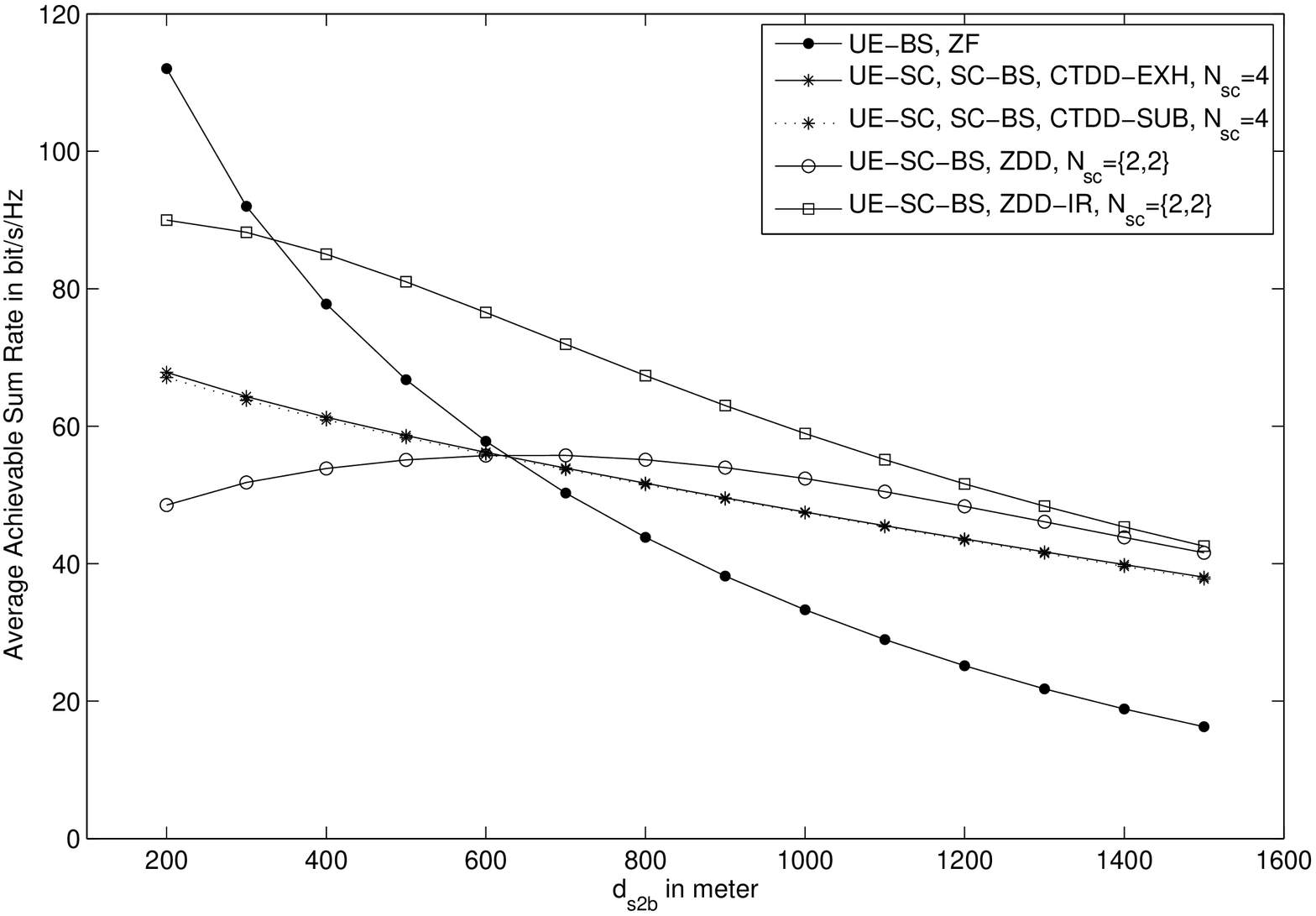}
\caption{Uplink average achievable sum rates of different techniques for $N_{\mathrm{sc}}=4$ and $N_{\mathrm{ue}}=2$ with different $d_{\mathrm{s2b}}$ values.}
\label{fig:ul_Nue2_Nsc4}
\end{figure}

For the uplink scenario, similarly to the downlink, the distance between each SC and the BS is at the range of $d_{\mathrm{s2b}} \in\{200\mathrm{m}, 300\mathrm{m}, \ldots, 1500\mathrm{m}\}$. The average achievable uplink sum-rate results in $\mathrm{bit/s/Hz}$ employing different techniques discussed in Section \ref{sec:strategies} are compared to the results of directly ZF decoding without SCs described in Section \ref{sec:basis}. Note that a larger $d_{\mathrm{s2b}}$ value implies worse SC-BS and UE-BS links. For $N_{\mathrm{ue}}=1$, Fig. \ref{fig:ul_Nue1_Nsc4} presents the average achievable sum-rate results of $N_{\mathrm{sc}}=4$. Similarly, in the case of $N_{\mathrm{ue}}=2$ and $N_{\mathrm{sc}}=4$, the results are shown in Fig. \ref{fig:ul_Nue2_Nsc4}.

Similarly to the downlink, compared to directly ZF decoding without SCs, CTDD-EXH with  multiple-antenna SCs and better UE-SC and SC-BS links could offer a significantly better average achievable sum rate, when the UE-BS links are not very strong, as shown in the figures. Moreover, it could be expected that a larger $N_{\mathrm{sc}}$ value could offer a better average achievable sum rate. Although both techniques achieve a lower average achievable sum rate as the SC-BS links become worse, directly ZF decoding decreases much quicker. Hence, if the UE-BS links are sufficiently strong, directly ZF decoding could offer the better average achievable sum rate as shown in the figures. However, it could be expected that by adding $N_{\mathrm{sc}}$, CTDD-EXH always has the potential to offer the better average achievable sum rate than directly ZF decoding, when the UE-SC links are better than the UE-BS links, which is mostly the practical case. Furthermore, the suboptimal CTDD-SUB could offer almost the same results as CTDD-EXH.

In the case of ZDD, the uplink average achievable sum rate increases to a point as $d_{\mathrm{s2b}}$ becomes larger, then it becomes relatively stable for a certain range before declining, as shown in the figures. The reason is that, as $d_{\mathrm{s2b}}$ increases, although the SC-BS links become weaker, the interference from the UE-SC links reduces as well with even a quicker rate. Therefore, the  SC-BS links generally become better as $d_{\mathrm{s2b}}$ enlarges, until the interference becomes lower than the noise power, from where the SC-BS links start to become weaker as $d_{\mathrm{s2b}}$ increases. On the other hand, the UE-SC links are independent of $d_{\mathrm{s2b}}$. As a result, the overall average achievable sum rate is at first limited by the SC-BS links so that it improves as $d_{\mathrm{s2b}}$ increases, until it becomes limited by the UE-SC links. From there, the overall average achievable sum rate becomes independent of $d_{\mathrm{s2b}}$ to a certain point before being limited by the SC-BS links again and starts to reduce. Due to the above properties, after a certain value of $d_{\mathrm{s2b}}$, ZDD is better than directly ZF decoding without SCs as shown in the figures. The comparison between ZDD and CTDD-EXH is more complicated. However, even with the conservative ZDD method, ZDD could offer the better average achievable sum rate than CTDD, as shown in the figures.

As for ZDD-IR, as the interference of the SC-BS links from the UE-SC links is rejected, it is generally better than ZDD as shown in the figures. For the range that the average achievable sum rate is limited by the UE-SC links, it is independent of $d_{\mathrm{b2s}}$ and is slightly better than ZDD as shown in Fig. \ref{fig:ul_Nue1_Nsc4}. The reason for the slightly better results is that ZDD-IR offers better SC-BS links than ZDD. On the other hand, for the range that the average achievable sum rate is limited by the SC-BS links, it is better than ZDD due to the improved SC-BS links, as shown in the figures. Moreover, in this case, ZDD-IR decreases as the SC-BS links become worse. Eventually, ZDD-IR becomes similar to ZDD as shown in the figures.

In summary, similarly to the downlink case, if SCs are located very close to the BS, directly ZF decoding between UEs and the BS is the best strategy, which means that applying SCs is not an efficient choice in this case. On the other hand, when SCs are located with certain distances to the BS, all CTDD, ZDD, and ZDD-IR could be better than directly ZF decoding without SCs. Specifically, ZDD-IR is the best strategy for the middle range, while ZDD is efficient for the far range. Note that for some cases, e.g., Fig. \ref{fig:ul_Nue1_Nsc4}, ZDD could be sufficiently good for both middle and far ranges.

\subsection{Residual Self-Interference} \label{subsec:results_rsi}
As shown in Section \ref{subsec:results_dl} and Section \ref{subsec:results_ul}, ZDD and ZDD-IR could achieve better throughput results compared to CTDD and directly ZF without SCs, for both the downlink and uplink. However, realistically, RSI that results from imperfect self-interference cancellation exists for ZDD and ZDD-IR. RSI results in performance degradation for the received signals. In our case, it weakens the BS-SC links for the downlink and the UE-SC links for the uplink. Note that in the sections \ref{subsec:results_dl} and \ref{subsec:results_ul}, $0\mathrm{dB}$ RSI is assumed to offer the throughput performance upper bound. In practice, based on our experiments for current ZDD techniques, RSI is around $2\mathrm{dB}$ for conservative ZDD where the transmitting and receiving antennas are separate, while near $5\mathrm{dB}$ for complete ZDD where each antenna both transmits and receives signals. In this subsection, RSI is included for (\ref{eq:capacity_dl_b2s_zdd_wt}), (\ref{eq:capacity_ul_u2s_zdd_wt}), (\ref{eq:capacity_dl_b2s_zdd-ir_wt}), and (\ref{eq:capacity_ul_u2s_zdd-ir_wt}), and simulation results for ZDD and ZDD-IR with RSI are provided while directly ZF without SCs and CTDD-EXH are shown as reference curves.

\begin{figure}[!t]
\centering \includegraphics[width = 1.0\linewidth]{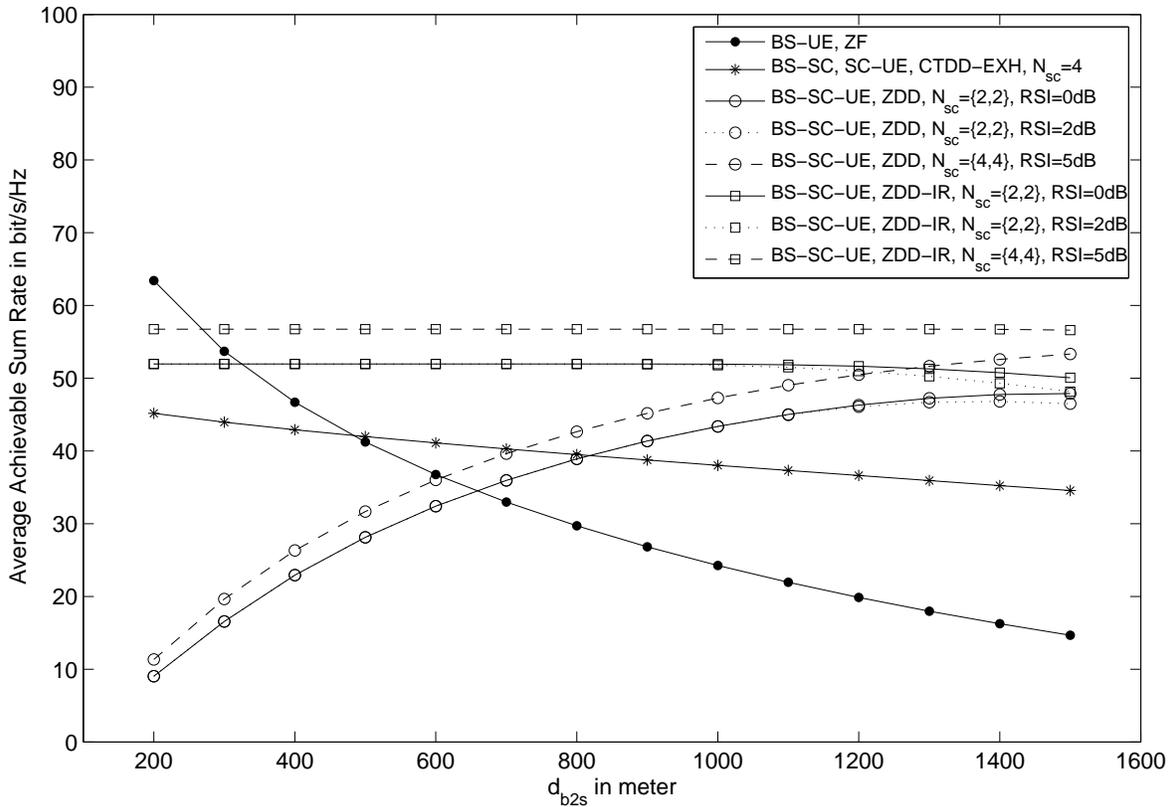}
\caption{Downlink average achievable sum rates of different RSI cases for $N_{\mathrm{sc}}=4$ and $N_{\mathrm{ue}}=1$ with different $d_{\mathrm{b2s}}$ values.}
\label{fig:dl_Nue1_Nsc4_rsi}
\end{figure}

\begin{figure}[!t]
\centering \includegraphics[width = 1.0\linewidth]{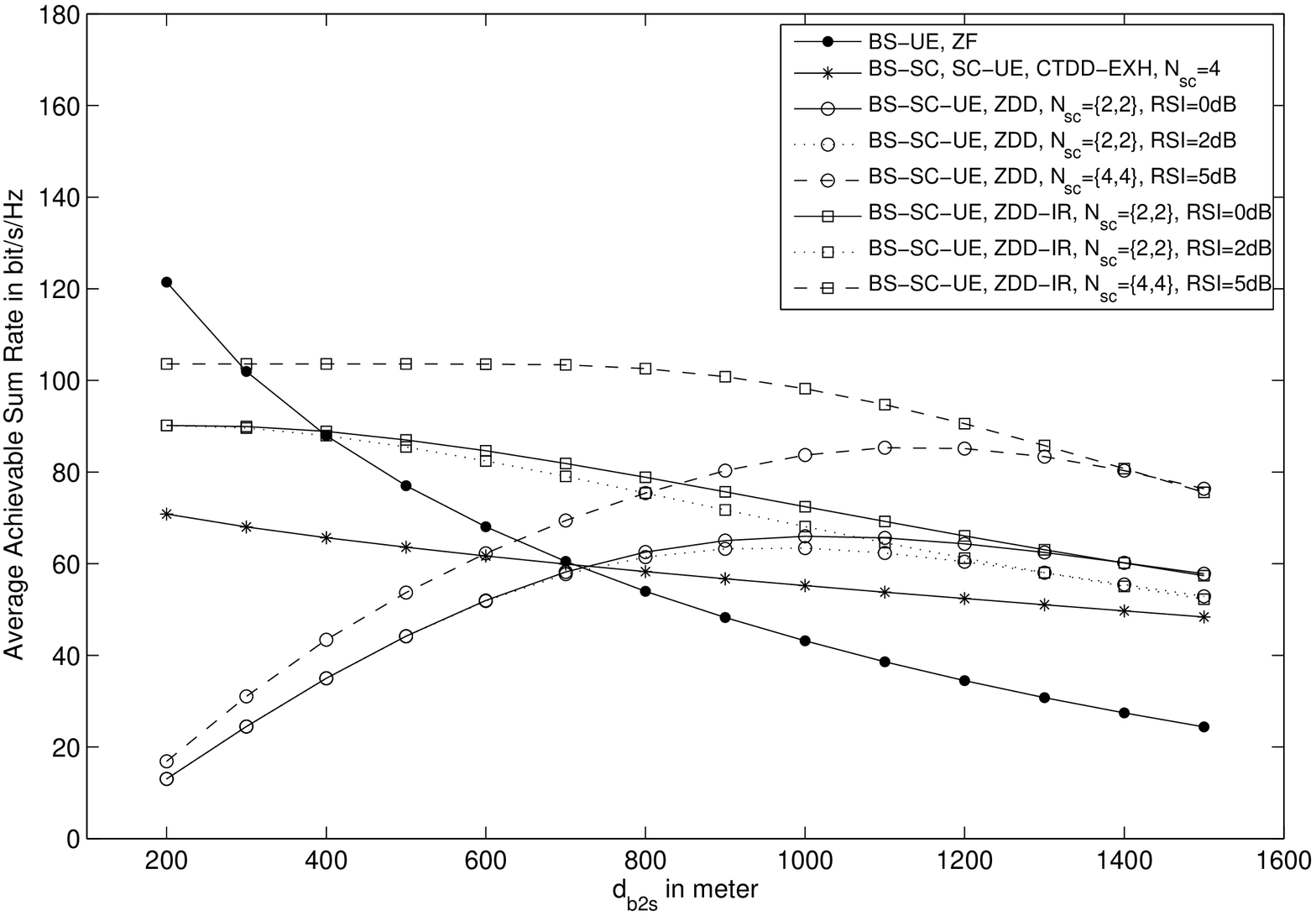}
\caption{Downlink average achievable sum rates of different RSI cases for $N_{\mathrm{sc}}=4$ and $N_{\mathrm{ue}}=2$ with different $d_{\mathrm{b2s}}$ values.}
\label{fig:dl_Nue2_Nsc4_rsi}
\end{figure}

Fig.\ref{fig:dl_Nue1_Nsc4_rsi} and Fig. \ref{fig:dl_Nue2_Nsc4_rsi} show the downlink average achievable sum-rate results of different RSI cases with $N_{\mathrm{sc}}=4$ for $N_{\mathrm{ue}}=1$ and $N_{\mathrm{ue}}=2$ respectively. In the case of conservative ZDD, since RSI only affects the BS-SC links, the average achievable sum-rate results in $\mathrm{bit/s/Hz}$ for ZDD and ZDD-IR are worse than the no RSI case when the BS-SC links limit the results, as shown in the figures. Despite of the losses caused by RSI, conservative ZDD and ZDD-IR could still outperform CTDD and direct ZF without SCs. For the more complicated complete ZDD, the number of spatial multiplexing streams is increased for the BS-SC links and the array gains are improved for the SC-UE links, compared to conservative ZDD. More streams of complete ZDD for the BS-SC links could provide greater throughput increase than the decrease caused by higher RSI. In addition, the SC-UE links of complete ZDD are better than conservative ZDD. Hence, even with $5\mathrm{dB}$ RSI, complete ZDD and ZDD-IR could achieve better results than conservative ZDD and ZDD-IR respectively, as shown in the figures. 

\begin{figure}[!t]
\centering \includegraphics[width = 1.0\linewidth]{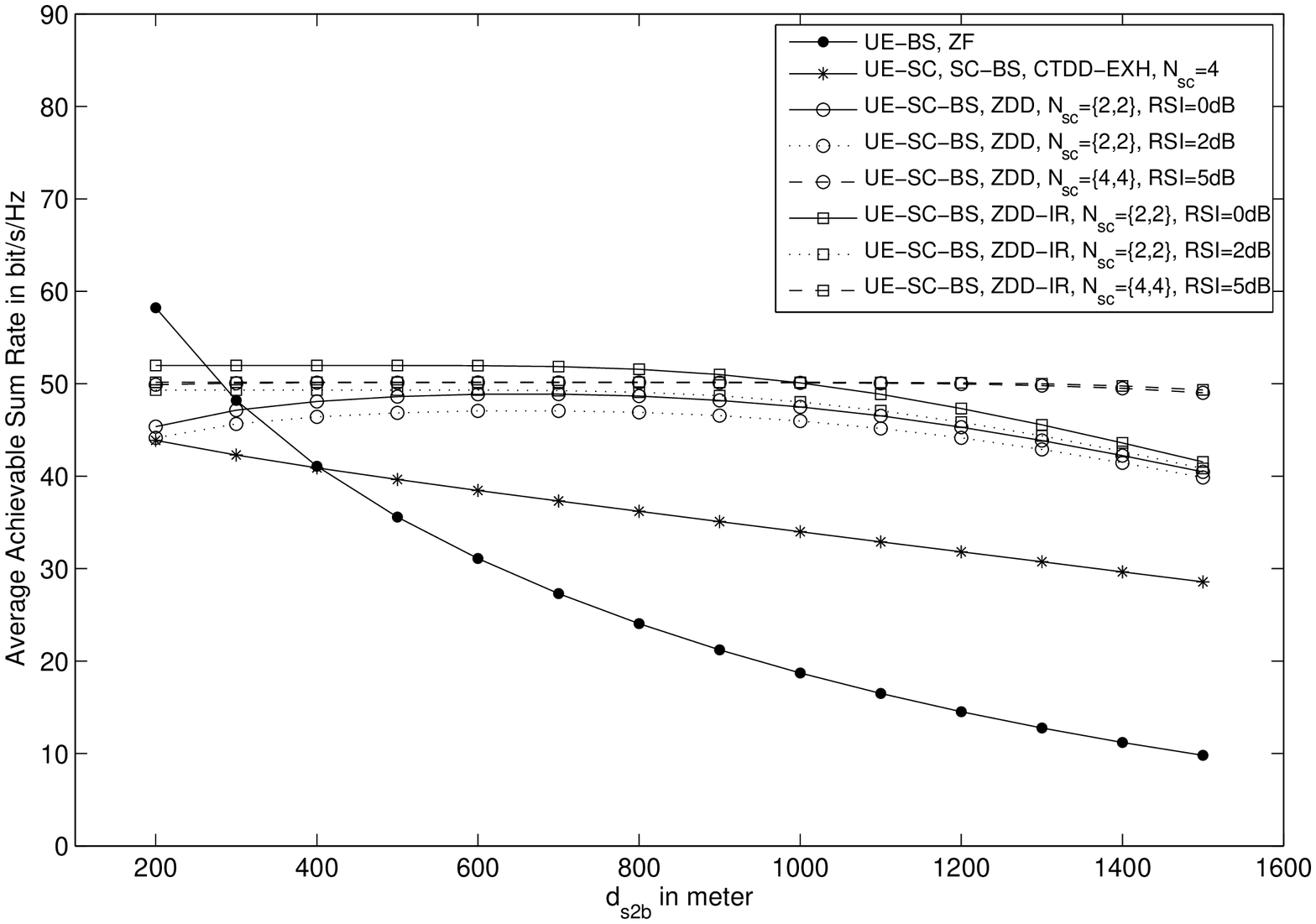}
\caption{Uplink average achievable sum rates of different RSI cases for $N_{\mathrm{sc}}=4$ and $N_{\mathrm{ue}}=1$ with different $d_{\mathrm{s2b}}$ values.}
\label{fig:ul_Nue1_Nsc4_rsi}
\end{figure}

\begin{figure}[!t]
\centering \includegraphics[width = 1.0\linewidth]{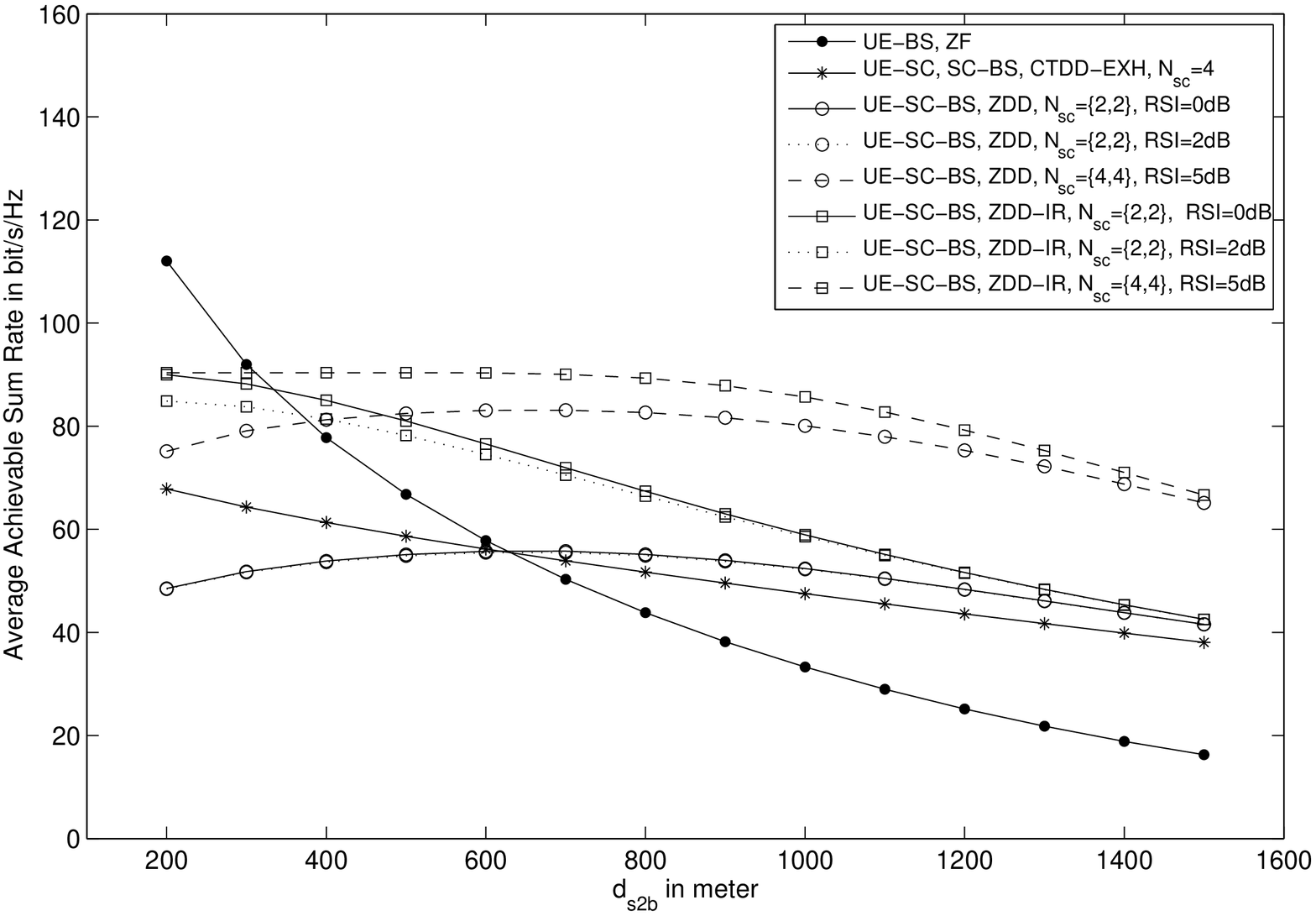}
\caption{Uplink average achievable sum rates of different RSI cases for $N_{\mathrm{sc}}=4$ and $N_{\mathrm{ue}}=2$ with different $d_{\mathrm{s2b}}$ values.}
\label{fig:ul_Nue2_Nsc4_rsi}
\end{figure}

Similarly, the uplink average achievable sum-rate results in $\mathrm{bit/s/Hz}$ of different RSI cases with $N_{\mathrm{sc}}=4$ for $N_{\mathrm{ue}}=1$ and $N_{\mathrm{ue}}=2$ are shown in Fig. \ref{fig:ul_Nue1_Nsc4_rsi} and Fig. \ref{fig:ul_Nue2_Nsc4_rsi} respectively. For the uplink, RSI only affects the UE-SC links. Therefore, compared to the no RSI case, the average achievable sum-rate results for ZDD and ZDD-IR of conservative ZDD are worse when the UE-SC links limit the results, as shown in the figures. Nevertheless, even with RSI, conservative ZDD and ZDD-IR could still outperform CTDD and direct ZF without SCs. Note that in Fig. \ref{fig:ul_Nue2_Nsc4_rsi}, conservative ZDD curves with and without RSI achieve almost the same average achievable sum rate because they are both limited by the SC-BS links. In the more complicated complete-ZDD case, compared to conservative ZDD, the number of spatial multiplexing streams is increased for the SC-BS links and the array gains are improved for the UE-SC links. The improved array gains of the UE-SC links could offer more throughput increase than the degradation due to worse RSI. Moreover, the increased streams of ZDD could improve the SC-BS links. As a result, complete ZDD and ZDD-IR, even with $5\mathrm{dB}$ RSI, could provide greater results than conservative ZDD and ZDD-IR respectively, as shown in the figures. Note that in Fig. \ref{fig:ul_Nue1_Nsc4_rsi}, complete ZDD and complete ZDD-IR achieve almost the same average achievable sum rate because they are both limited by the UE-SC links.

\subsection{Other Related Discussions} \label{subsec:discussions}
Two frequency bands are assumed in this paper where one is used for the data transmission of the BS while the other one is used for the data transmission of SCs, as shown in Fig. \ref{fig:eg_inband}. Since the main purpose of this paper is to investigate the strategies and throughput gains by using massive MIMO as the SC wireless backhaul, only the band used for data transmission of SCs is discussed. The proper applications of SC in-band wireless backhaul in heterogeneous networks and the comparisons with other network schemes are considered as future works. Note that the assumption of two frequency bands might require major changes to current wireless networks. However, in order to achieve the extremely high requirements for the next-generation wireless systems, e.g., 5G and beyond, new technologies are needed and have been considered, e.g., millimeter-wave systems \cite{Rappaport_mmWave,Rappaport_mmWave_Networks}. Hence, major changes are possible and may happen in the near future. 

In this paper, all SCs are assumed to be equipped with wireless backhaul
. Note that a massive MIMO wireless backhaul station is not necessarily placed at a BS. In addition, all SCs do not necessarily use wireless in-band backhaul in a network. In other words, SCs with wired backhaul can still use the frequency band employed for the data transmission of the BS, e.g., $F_1$ in Fig. \ref{fig:eg_inband}, so that they do not interfere SCs with the proposed wireless backhaul using the other frequency band, e.g., $F_2$ in Fig. \ref{fig:eg_inband}, and vice versa. 
Because SCs with wireless in-band backhaul can be specifically deployed to the locations where SCs are needed but without wired backhaul, it is possible that their deployment density is much sparser than SCs with wired backhaul. Moreover, for a dense SC network, the proposed massive MIMO in-band wireless backhaul can be employed to replace the wired backhaul of some carefully selected SCs to alleviate the inter-SC interference taking advantage of the second band.   

All SCs with the proposed in-band wireless backhaul are assumed to use the same band, e.g., $F_2$ in Fig. \ref{fig:eg_inband}, in this paper. As mentioned in Section \ref{sec:strategies}, SCs are assumed to be carefully located so that inter-SC interference can be neglected, which is possible. As shown in Fig. \ref{fig:eg_inband}, since two bands are assumed, 
SCs can be located at where they are needed rather than to cover the whole area. Strong inter-SC interference can be avoided by careful location selection. Even if strong inter-SC interference exists in two neighboring SCs, advanced interference management schemes included in LTE-A such as ICIC and Coordinated Multipoint (CoMP) can be applied to address this issue \cite{Tran_LTE-A}. Note that the main goal of this paper is to investigate the strategies and throughput improvement by using massive MIMO as the SC wireless backhaul, rather than to address the inter-cell interference issue. 

In this paper, it is assumed that all time-frequency resources of a SC are allocated to a single UE. When multiple UEs are served by a SC, they can be allocated different time-frequency resources. As a result, the overall system with multiple UEs per SC can be considered as multiple separate systems with single UE per SC. Then, the analysis in Section \ref{sec:strategies} can be easily generalized to the case of multiple UEs per SC. When the simulated number of drops, where a drop means a single statistical event, with single UE per SC is sufficiently large as in our simulations, it equals to the simulation of multiple UEs per SC, because the average achievable sum rate over the serving area of a SC for each UE in terms of $\mathrm{bit/s/Hz}$ is the same. Therefore, the simulation results based on single UE per SC in this section 
in fact can reflect the practical case where multiple UEs are served per SC. 

The interference out of the area of interest, e.g., from other macro cells, is not considered in this paper. However, we believe that it would not undermine the main goal of this paper, i.e., to investigate the strategies and throughput improvement by using massive MIMO as the SC wireless backhaul. Note that the interference out of the area of interest mainly affects cell edge where the average achievable rate is already relatively small. In addition, the probability of a UE located at cell edge is generally smaller than not located at cell edge. As a result, even if the average achievable rate of cell edge is substantially reduced by it, the average achievable sum rate over the whole considered area as presented in Fig. \ref{fig:system} might only slightly decrease. Since the interference out of the area of interest would introduce a similar level of interference, which could be considered as additional noise, to all investigated methods in this paper, it can be expected that the relative relations and the associated conclusions presented in this section still hold hence are practically relevant, even if the absolute values might be slightly lower.

In this paper, uncorrelated Rayleigh fading is applied as the channel model and perfect channel state information is assumed for simulations to initially investigate whether a massive MIMO wireless in-band backhaul station supporting multiple SCs is a valid application. Because the results are promising, simulations with more practical assumptions such as more realistic channel models and channel estimation errors are considered as future works.

\section{Conclusions} \label{sec:conclusions}
In this paper, three strategies of SC in-band wireless backhaul in massive MIMO systems were provided and discussed. When the links between the BS and UEs are not very good, CTDD, ZDD, and ZDD-IR could achieve significant throughput improvement compared to directly ZF beamforming without SCs. Among the three strategies, CTDD is the simplest one and could achieve a decent throughput gain. ZDD requires the capability of self-interference cancellation at SCs and could achieve better throughput than CTDD depending on conditions, even with RSI. Other than self-interference cancellation, ZDD-IR requires the additional interference rejection process at the BS, but the increased complexity could result in generally better throughput than CTDD and ZDD. In summary, SC in-band wireless backhaul has the potential to increase the throughput for massive MIMO systems. Its proper applications in heterogeneous networks, their comparisons with other network schemes, and simulations with more practical assumptions are considered as future works. 
\section*{Acknowledgment}
The authors would like to thank the anonymous reviewers whose valuable comments improved the quality of the paper.

\bibliographystyle{IEEEtran}
\bibliography{IEEEabrv,Mybib}
\end{document}